\documentclass[12pt,english]{article}
\pdfoutput=1
\usepackage{graphicx,array}
\usepackage{cite}
\usepackage{color}
\usepackage{latexsym}
\usepackage{amsthm}
\usepackage{amsmath}
\usepackage[titletoc]{appendix}
\usepackage{enumitem}
\usepackage{amssymb}
\usepackage{hyperref}
\usepackage[hang,flushmargin]{footmisc} 
\usepackage{stmaryrd}
\numberwithin{equation}{section}
\def\simgt{\mathrel{\lower2.5pt\vbox{\lineskip=0pt\baselineskip=0pt
           \hbox{$>$}\hbox{$\sim$}}}}
\def\simlt{\mathrel{\lower2.5pt\vbox{\lineskip=0pt\baselineskip=0pt
           \hbox{$<$}\hbox{$\sim$}}}}

\newcommand{\nn}{\nonumber}
\newcommand{\be}{\begin{equation}}
\newcommand{\ee}{\end{equation}}
\newcommand{\bea}{\begin{eqnarray}}
\newcommand{\eea}{\end{eqnarray}}
\newcommand{\Eq}[1]{Eq.~(\ref{#1})}
\newcommand{\Eqs}[2]{Eqs.~(\ref{#1}) and (\ref{#2})}
\newcommand{\Sec}[1]{Sec.~\ref{#1}}
\newcommand{\Secs}[2]{Secs.~\ref{#1} and \ref{#2}}
\newcommand{\Fig}[1]{Fig.~\ref{#1}}
\newcommand{\App}[1]{App.~\ref{#1}}

\newcommand{\Ref}[1]{Ref.~\cite{#1}}
\newcommand{\Refs}[2]{Refs.~\cite{#1} and \cite{#2}}

\newcommand{\mPl}{m_{\rm Pl}}

\newcommand*\oline[1]{%
  \vbox{%
    \hrule height 0.5pt
    \kern0.68ex
    \hbox{%
      \kern-0.1em
      \ifmmode#1\else\ensuremath{#1}\fi
      \kern-0.1em
    }
  }
}

\definecolor{nicered}{rgb}{0.7,0.1,0.1}
\definecolor{nicegreen}{rgb}{0.1,0.5,0.1}
\hypersetup{
 colorlinks,citecolor=black,linkcolor=black,urlcolor=black}

\begin{document}
\interfootnotelinepenalty=10000
\baselineskip=18pt
\hfill CALT-TH-2014-146
\hfill

\vspace{1.5cm}
\thispagestyle{empty}
\begin{center}
{\LARGE\bf
Infrared Consistency and \\ \medskip  the Weak Gravity Conjecture
}\\
\bigskip\vspace{1.5cm}{
{\large Clifford Cheung and Grant N. Remmen}
} \\[7mm]
 {\it Walter Burke Institute for Theoretical Physics, \\[-1mm]
    California Institute of Technology, Pasadena, CA 91125}\let\thefootnote\relax\footnote{e-mail: \url{clifford.cheung@caltech.edu}, \url{gremmen@theory.caltech.edu}} \\
 \end{center}
\bigskip
\centerline{\large\bf Abstract}

\begin{quote} \small
The weak gravity conjecture (WGC) asserts that an Abelian gauge theory coupled to gravity is  inconsistent unless it contains a particle of charge $q$ and mass $m$ such that $q \geq m/\mPl$.  This criterion is obeyed by all known ultraviolet completions and is needed to evade pathologies from stable black hole remnants.  In this paper, we explore the WGC from the perspective of low-energy effective field theory.  Below the charged particle threshold, the effective action describes a photon and graviton interacting via higher-dimension operators.  We derive infrared consistency conditions on the parameters of the effective action using {\it i}) analyticity of light-by-light scattering, {\it ii}) unitarity of the dynamics of an arbitrary ultraviolet completion, and {\it iii}) absence of superluminality and causality violation in certain non-trivial backgrounds.  For convenience, we begin our analysis in three spacetime dimensions, where gravity is non-dynamical but has a physical effect on photon-photon interactions.  We then consider four dimensions, where propagating gravity substantially complicates all of our arguments, but bounds can still be derived.   Operators in the effective action arise from two types of diagrams: those that involve electromagnetic interactions (parameterized by a charge-to-mass ratio $q/m$) and those that do not (parameterized by a coefficient $\gamma$). Infrared consistency implies that $q/m$ is bounded from below for small $\gamma$.
\end{quote}

\setcounter{footnote}{0}

\newpage
\tableofcontents

\newpage

\section{Introduction}
\label{sec:intro}

The weak gravity conjecture (WGC) \cite{WGC} asserts a powerful restriction on any Abelian gauge theory coupled consistently to gravity.  In particular, it mandates the existence of a state of charge $q$ and mass $m$ satisfying\footnote{Throughout, we use natural units for mass and charge in which $4\pi G =\epsilon_0=1$, with $(+, -, -, \ldots)$ metric signature and curvature tensors $R_{\mu\nu} = R^\rho_{\;\; \mu \rho \nu}$ and $R^\rho_{\;\; \mu \sigma \nu} = \partial_{\sigma} \Gamma^\rho_{\;\; \mu \nu} - \partial_\nu \Gamma^\rho_{\;\; \mu\sigma}  + \Gamma^\rho_{\;\; \alpha \sigma }\Gamma^\alpha_{\;\;\mu \nu} - \Gamma^\rho_{\;\; \alpha\nu} \Gamma^\alpha_{\;\; \mu\sigma}$, all for arbitrary spacetime dimension $D$.}
\be 
q  \geq  m .
\label{eq:WGC}
\ee
Informally, the WGC states that ``gravity is the weakest force'' because it bounds the gravitational charge of a state from above by its electric charge.  The WGC is a beautiful and sharply defined criterion demarcating the landscape from the swampland.

The authors of \Ref{WGC} supported their conjecture with numerous examples from field theory and string theory, all satisfying the WGC.  Moreover, they offered an elegant argument by contradiction in favor of the WGC.  By conservation of charge and energy, the state with the largest charge-to-mass ratio cannot decay, so violation of the WGC implies the absolute stability of extremal black holes, which exactly saturate \Eq{eq:WGC}.  However, stable black hole remnants are thought to be pathological \cite{SusskindRemnants,tHooft,Bousso,GiddingsRemnant}, so the authors of \Ref{WGC} argued that the WGC is mandatory in any theory with an Abelian gauge symmetry.

 In this paper, we explore the WGC from the viewpoint of effective field theory.  Our central question is simple: does violation of the WGC induce a pathology in the infrared?  To seek an answer, we consider energies far below the charged particle threshold, where the dynamics are described by  photons and gravitons interacting via higher-dimension operators:
 \be
\begin{aligned}
\mathcal{L}  
&= - \frac{1}{4}F_{\mu\nu} F^{\mu\nu}  - \frac{1}{4} R \\
&\hspace{4.7mm}+a_1  (F_{\mu\nu} F^{\mu\nu})^2 + a_2  ( F_{\mu\nu} \tilde F^{\mu\nu})^2 \\
&\hspace{4.7mm}+  b_1 F_{\mu\nu} F^{\mu\nu} R +  b_2 F_{\mu\rho} F_{\nu}^{\;\;\rho} R^{\mu \nu}+b_3 F_{\mu\nu} F_{\rho\sigma} R^{\mu\nu\rho\sigma} \\
&\hspace{4.7mm}+ c_1 R^2 + c_2 R_{\mu\nu}R^{\mu\nu} +c_3  R_{\mu\nu\rho\sigma} R^{\mu\nu\rho\sigma},
\end{aligned}
\label{eq:EH}
\ee
where $\tilde{F}_{\mu\nu} = \epsilon_{\mu\nu\rho\sigma}F^{\rho\sigma}/2$.  We have dropped terms like $(\nabla_\mu F_{\nu\rho})^2$ and $(\nabla_\mu F^{\mu\nu})^2$, which in the absence of charged sources can be written in terms of the operators already included.

Electromagnetic interactions induce contributions to $a_i$ and $b_i$ that depend on the charges and masses of every state in the spectrum.  Each contribution grows with charge and scales inversely with mass, so they are dominated by the state in the spectrum with the largest charge-to-mass ratio, which we will write as $z =q/m$.  Crucially, the operator coefficients in the effective Lagrangian \eqref{eq:EH} are sensitive to the same quantity as the WGC, which posits that
\be
z\geq 1.\label{eq:WGCz} 
\ee
Because the photon-graviton effective action is $z$-dependent, there is hope that an analysis of the infrared dynamics might shed light on the WGC.

From a purely low-energy perspective, it would seem reasonable for the landscape of high-energy completions to span all values of the parameters in the effective action.  However, as discussed in \Ref{IRUV}, this is a misconception: some effective theories are intrinsically pathological and never emerge from consistent ultraviolet dynamics.  This occurs, for example, in the Euler--Heisenberg Lagrangian \cite{EulerHeisenberg,Schwinger,Weisskopf}, which is \Eq{eq:EH} in the limit that gravity is decoupled.  When $a_i < 0$, the theory admits superluminal photon propagation  and non-analyticity in the light-by-light scattering amplitude.  Unsurprisingly, $a_i\geq 0$ in all known ultraviolet completions. More recently, bounds on graviton interactions were derived in \Ref{graviton3}.

The purpose of this paper is to apply similar methods to determine infrared consistency conditions on the effective action describing the low-energy interactions of photons and gravitons.  In particular we derive constraints on the parameters of \Eq{eq:EH} from three independent criteria:
\begin{itemize}

\item[{\it i})] {\bf Analyticity.}  We study the analytic properties of the light-by-light scattering amplitude.  Forward dispersion relations constrain the effective theory parameters.

\item[{\it ii})] {\bf  Unitarity.}  We construct a spectral representation parameterizing an arbitrary ultraviolet completion.    Forbidding ghosts and tachyons constrains the effective theory parameters.

\item[{\it iii})] {\bf Causality.}  We compute the speed of light in certain non-trivial backgrounds.  Absence of superluminality and causality violation constrains the effective theory parameters.

\end{itemize}
As a warmup, we study the photon-graviton effective theory in three spacetime dimensions (3D), where gravity is purely topological \cite{3DGRStan}.  While the graviton is non-propagating, it still mediates contact interactions for the photon.   Remarkably, arguments from analyticity, unitarity, and causality all imply an identical constraint on the parameters of the effective theory:
\be 
a' \geq 0,
\label{eq:introbound3d}
\ee
where $a'= a_1 + b_1 - b_3 + c_1  + c_2 + 3 c_3 $.  We can, however, learn more by inputting additional assumptions about the ultraviolet completion.  For example, consider the case where the dominant contributions to $a_i$ and $b_i$ originate from diagrams involving  electromagnetic interactions of a fermion with charge-to-mass ratio $z$. As we will see, \Eq{eq:introbound3d} then implies a constraint on a two-dimensional parameter space spanned by $z$ and a coefficient $\gamma$ parameterizing purely gravitational corrections to the effective action.  The theory automatically satisfies our consistency conditions if $\gamma$ exceeds a certain critical value. 
 However, below this critical value, the theory is consistent only for certain values of $z$.  In particular, for small $\gamma$, infrared consistency implies that $z \geq 1$, a 3D version of the WGC.

Subsequently, we move on to four spacetime dimensions (4D), where dynamical gravity introduces a litany of subtleties, which we discuss at length in the body of the paper.  For now, let us simply summarize our results.  As we will see, unitarity arguments imply that
\be 
 a_1' \geq 0   \qquad{\rm and}\qquad
a_2' \geq 0,
\label{eq:introbound4d2}
\ee
where $a_1' = a_1- b_2/2  -b_3 +c_2+ 4c_3$ and $a_2' = a_2- b_2/2 -b_3 +c_2+ 4c_3$.  Meanwhile, 
 the absence of superluminal photon propagation in certain non-trivial backgrounds implies that
\be 
a_1' + a_2' \geq 0,
\label{eq:introbound4d1}
\ee
which is also a consequence of the unitarity bounds in \Eq{eq:introbound4d2}.  Analyticity arguments, on the other hand, are suspect in 4D because they rely  crucially on the forward light-by-light scattering amplitude, which is ill-defined due to singular $t$-channel graviton exchange \cite{IRUV}.  Nevertheless, if one can assume that dispersion relations apply to contributions to the forward amplitude from higher-dimension operators,  then remarkably,
 \Eq{eq:introbound4d1} can also be derived as a consequence of analyticity.  In this sense, arguments from analyticity, unitarity, and causality in 4D all point to the set of mutually consistent bounds in \Eqs{eq:introbound4d2}{eq:introbound4d1}.

These bounds imply 4D constraints on the parameter space defined by $z$ and the coefficients $\gamma$ that parameterize purely gravitational effects.  Our results in 4D are summarized  in \Fig{fig:bound4d}.  In all cases, when $\gamma$ is small, infrared consistency implies a lower bound on $z$ that is numerically stronger than the WGC.  Curiously, in this regime we find that \Eq{eq:introbound4d1} results in the exact same bound for fermions and scalars: $z\geq 2$.

The remainder of our paper is structured as follows.  In \Sec{sec:3d}, we 
derive constraints on the photon-graviton effective action in 3D coming from analyticity, unitarity, and causality.   We then present the analogous arguments for the photon-graviton effective action in 4D in \Sec{sec:4d}.  Finally, we conclude and discuss future directions in \Sec{sec:conclusions}.

\section{Three Dimensions}

\label{sec:3d}

\subsection{Setup and Bounds (3D)}

\label{sec:eff3d}

To begin, we re-express \Eq{eq:EH} in a form convenient for studying the dynamics of interacting photons.  Specifically, we eliminate all dependence on the spacetime curvature in favor of the electromagnetic field strength.  We start by rewriting the Riemann tensor in terms of the  Ricci scalar, Ricci tensor, and Weyl tensor, which in $D$ dimensions is\footnote{Throughout, square brackets denote un-normalized antisymmetrization, {\it viz}. $T_{[\mu\nu]} = T_{\mu\nu} - T_{\nu\mu}$.}
\be 
C_{\mu\nu\rho\sigma} = R_{\mu\nu\rho\sigma} - \frac{1}{D-2} (g_{\mu[\rho} R_{\sigma] \nu}-g_{\nu[\rho}R_{\sigma] \mu}) +\frac{1}{(D-1)(D-2)} R g_{\mu[\rho}g_{\sigma]\nu},
\label{eq:Weyldef}
\ee
where in 3D the Weyl tensor identically vanishes and \Eq{eq:Weyldef} implies that
\be 
C_{\mu\nu\rho\sigma}C^{\mu\nu\rho\sigma} = R^2 - 4 R_{\mu\nu} R^{\mu\nu} + R_{\mu\nu\rho\sigma}R^{\mu\nu\rho\sigma},
\ee
so the Gauss--Bonnet term vanishes identically in 3D.
Next, we eliminate all dependence on the Ricci tensor and Ricci scalar in the higher-dimension operators by rewriting them via the tree-level Einstein field equations,
\be 
R_{\mu\nu} - \frac{1}{2} g_{\mu\nu}R = 2 T_{\mu\nu}, 
\label{eq:Einstein}
\ee 
which at the order of the Lagrangian \eqref{eq:EH} is equivalent to a field redefinition of the graviton.  Meanwhile, the energy-momentum tensor is
\be 
T_{\mu\nu} = - F_{\mu\rho}F_\nu^{\;\; \rho} +\frac{1}{4}  g_{\mu\nu} F_{\rho\sigma} F^{\rho \sigma},
\label{eq:T}
\ee 
so \Eq{eq:EH} can be expressed solely in terms of the electromagnetic field strength. 
In particular, \Eqs{eq:Einstein}{eq:T} imply that $R^2 = R_{\mu\nu}R^{\mu\nu} = (F_{\mu\nu}F^{\mu\nu})^2$.

At leading order in derivatives, the only invariants constructed from the electromagnetic field strength are $(F_{\mu\nu}F^{\mu\nu})^2$ and $F_{\mu\rho} F_{\nu}^{\;\;\rho}F^{\mu}_{\;\;\sigma} F^{\nu \sigma}$.  In 3D, these are algebraically related by $ F_{\mu\rho} F_{\nu}^{\;\;\rho}F^{\mu}_{\;\;\sigma} F^{\nu \sigma}= (F_{\mu\nu}F^{\mu\nu})^2/2$.  
Thus, the final form of the photon-graviton effective Lagrangian in 3D is remarkably simple:
\be 
 {\cal L} =  - \frac{1}{4}F_{\mu\nu} F^{\mu\nu}- \frac{1}{4} R 
 +  a' (F_{\mu\nu} F^{\mu\nu})^2 .
\label{eq:EH3d}
\ee
Here we have defined a new higher-dimension operator coefficient,
\be 
a' = a_1 + b_1 - b_3 + c_1  + c_2 + 3 c_3,
\label{eq:WGCcoeffs}
\ee
written in terms of the original parameters in the Lagrangian \eqref{eq:EH} after discarding the operator $(F_{\mu\nu} \tilde F^{\mu\nu})^2$, which does not exist in 3D.

Next, we exploit a nice feature of 3D, namely, that a photon is equivalent to a scalar.  To simplify our calculations, we dualize the photon according to
\be 
F_{\mu\nu} =  i\epsilon_{\mu\nu\rho} \partial^{\rho} \phi,
\label{eq:dualize}
\ee 
where $\epsilon_{\mu\nu\rho}$ is the 3D Levi-Civita tensor and the overall coefficient is fixed so that $\phi$ is a canonically normalized state with positive norm.   After dualization, \Eq{eq:EH3d} becomes
\be 
{\cal L} =  \frac{1}{2} \partial_\mu \phi \partial^\mu \phi - \frac{1}{4} R 
+ 4 a'  (\partial_\mu \phi \partial^\mu \phi)^2,
\label{eq:scalarL}
\ee 
which is our final form for the photon-graviton effective Lagrangian in 3D.  The underlying gauge symmetry of the photon is encoded in the shift symmetry of $\phi$.

\begin{figure}[t] 
\centerline{\includegraphics[width=.85\columnwidth]{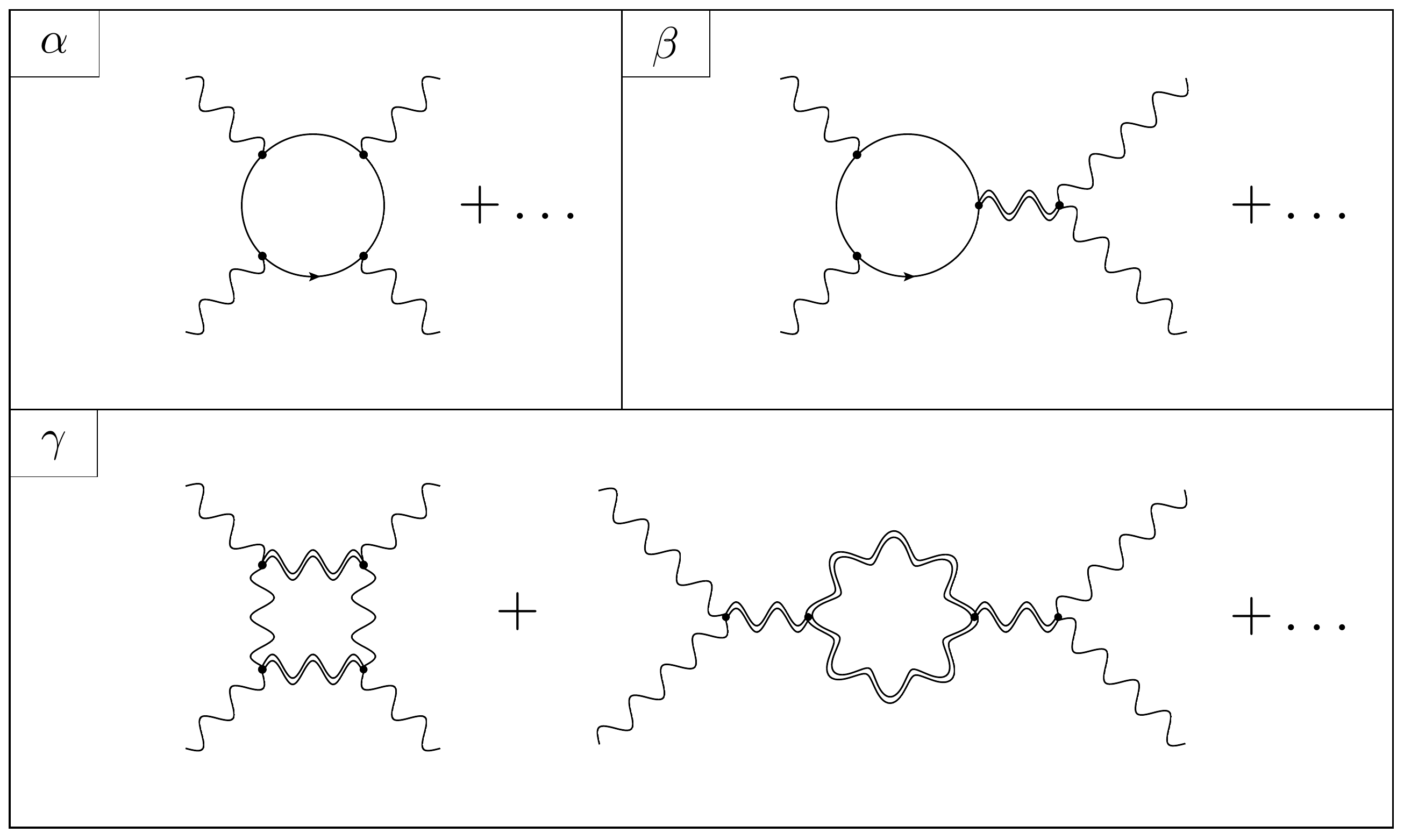}}
\caption{Diagrams involving photons (single wavy), gravitons (double wavy), and charged matter (solid) that contribute to light-by-light scattering, organized in terms of scaling with $z=q/m$, as defined in \Eq{eq:gamma}.  Here, $\gamma$ parameterizes purely gravitational corrections.  }\label{fig:abc}
\end{figure}

As we will derive shortly, the constraints from analyticity, causality, and unitarity in 3D all imply the exact same constraint, 
\be 
a' \geq 0.
\label{eq:consistency3d}
\ee
How might this bound constrain the spectrum of the ultraviolet completion? As noted earlier, the coefficients $a_i$ and $b_i$ in \Eq{eq:EH} receive calculable contributions from every charged particle in the spectrum, but they are dominated by the state with the largest charge-to-mass ratio, defined to be $z=q/m$.  
Without loss of generality, we can thus expand $a'$ in powers of $z$ as
\be 
a' = \alpha z^4 + \beta z^2 + \gamma.
\label{eq:gamma}
\ee 
Primordially, $\alpha$, $\beta$, and $\gamma$ arise from diagrams with four, two, and zero insertions of the electromagnetic coupling, respectively, as shown in \Fig{fig:abc}.

By definition, $\alpha$ and $\beta$ are contributions coming from diagrams that contain  electromagnetic interactions.  For example, integrating out a charged fermion in 3D yields calculable threshold corrections to the higher-dimension operator coefficients \cite{Ritz,D&H},
\be 
a_1 = \frac{q^4}{1920 \pi m^5 }  \qquad{\rm and}\qquad 
(b_1,b_2,b_3) = \left(-\frac{q^2}{1152 \pi m^3 },  \;\frac{13q^2}{2880 \pi m^3 }, \;
 -\frac{q^2}{2880 \pi m^3 } \right). 
 \label{eq:abc3}
\ee
In 3D, $q$ has mass dimension $1/2$.  By substituting
\Eq{eq:abc3} into \Eq{eq:WGCcoeffs} and comparing to \Eq{eq:gamma}, we straightforwardly obtain $\alpha$ and $\beta$. Despite the complicated numerical factors in \Eq{eq:abc3}, we find that $\alpha/\beta = -1$.  Meanwhile, since $\gamma$ is independent of $q$,  it necessarily parameterizes all contributions arising solely from gravitational interactions.  These include the combination of coefficients $c_1 + c_2 + 3c_3$ in \Eq{eq:WGCcoeffs}.  Because $\gamma$ is incalculable within the low-energy effective theory, it should be thought of as a high-energy boundary condition encoding the gravitational dynamics of the ultraviolet completion.  Finally, rewriting \Eq{eq:consistency3d} in terms of $z$ and $\gamma$, we find that
\be 
z^2(z^2-1)   \geq -\gamma m \times 1920 \pi  .\label{eq:3Dcondition}
\ee
If $\gamma \geq 1/7680 \pi m$, 
then this bound is satisfied for any value of $z$. This is a sufficient albeit not necessary condition for satisfying bounds from analyticity, unitarity, and causality.  

On the other hand, it is interesting to consider the case in which the gravitational corrections are small, so $\gamma \sim 0$.   In this case, our bounds imply that 
\be 
z \geq 1,
\ee
which is the 3D analogue of the WGC in \Eq{eq:WGC}.   This result is interesting because the argument for \Eq{eq:WGC} from \Ref{WGC} derives from pathologies of stable extremal black holes, which do not exist in asymptotically-flat 3D spacetime.   In this sense,  infrared consistency conditions have more general applicability than the extremal black hole arguments of \Ref{WGC}.

{\it A priori}, the 3D effective theory could arise from the compactification of a higher-dimensional theory.  
  Of course, even then, the infrared consistency condition in \Eq{eq:consistency3d} would hold.  However, if the compactification scale were less than $m/q$, then interactions generated from integrating out the radion and the Kaluza--Klein modes would dominate over  those generated by the charged states.  In this case, $z$ would contribute negligibly to the effective action and infrared consistency would simply bound the parameter $\gamma$.

\subsection{Analyticity (3D)}\label{sec:Analyticity3}

In this section, we exploit the analytic properties of the light-by-light scattering amplitude to constrain the 3D effective Lagrangian in \Eq{eq:scalarL}.  Following the procedure of \Ref{IRUV}, we consider the scattering amplitude
\be 
{\cal M}(s,t) = 8a' (s^2 +t^2 + u^2),
\label{eq:M4}
\ee
where the Mandelstam variables satisfy $s+t+u =0$. The forward scattering amplitude is then ${\cal M}(s) = {\cal M}(s , t\rightarrow 0) = 16 a' s^2$.  Next, to extract the operator coefficient we compute the contour integral of ${\cal M}(s)/s^3$ around a contour ${\cal C}$ encircling the origin:
\be 
\begin{aligned}
16a' &= \oint_{\cal C} \frac{{\rm d}s}{2\pi i} \frac{ {\cal M}(s)}{s^3} =\oint_{{\cal C}'} \frac{{\rm d}s}{2\pi i} \frac{ {\cal M}(s)}{s^3} \\
&= \left(\int_{-\infty}^{-s_0} +\int_{s_0}^{\infty} \right)  \frac{{\rm d}s }{2\pi i} \frac{ \textrm{Disc}[{\cal M}(s)]}{s^3} + \textrm{boundary integral}.
\end{aligned}
\ee
Following \Ref{IRUV}, we have used the Cauchy integral theorem to deform ${\cal C}$ into a new contour ${\cal C}'$ composed of lines running just above and below the real axis plus a large circular boundary contribution at infinity.   The discontinuity function is
\be 
\textrm{Disc}[{\cal M}(s)] = {\cal M}(s +i\epsilon) - {\cal M}(s-i\epsilon) 
= 2i \textrm{Im} [{\cal M}(s)],
\ee
where the difference of terms arises from the contour integration above and below the real axis and we used analyticity of $\mathcal{M}(s)$ to apply the Schwarz reflection principle, $\mathcal{M}(s^*) = \mathcal{M}(s)^*$.  Deforming the contour is mathematically permitted, provided ${\cal M}(s)$ is analytic in the bulk of the complex $s$ plane and in the neighborhood of $s=0$.  The former is guaranteed by the usual stipulation that all non-analyticities of the S-matrix, {\it e.g.}, poles and branch cuts, occur near the real axis.  The latter is ensured by an additional physical input, which is that the scattering amplitude does not have branch cuts on the real axis extending to $s=0$.  At one-loop order in the effective action, light-by-light scattering will include massless branch cuts from a photon loop and two insertions of the $(F_{\mu\nu}F^{\mu\nu})^2$ operator.   
However, as discussed in \Ref{IRUV}, such cuts can be avoided by a slight deformation of the contour after introducing a regulator mass for the photon. Moreover, there are no branch cuts from gravitons, which are non-dynamical in 3D.
For concreteness, we define $s_0$ to be the mass squared of the lowest-lying degree of freedom  produced from light-by-light scattering, so ${\cal M}(s)$ is analytic in the region $|s| < s_0$.

The contour integral over ${\cal C}'$ includes a contribution from the discontinuity across the real axis as well as a contribution from infinity.  In $D$ dimensions, unitarity and polynomial boundedness of amplitudes implies the Froissart bound for large $|s|$, $|{\cal M}(s) | \lesssim |s \log^{D-2} s|$ \cite{Chaichian1,Chaichian2}, so  the boundary term is zero.  Evaluating the contour integral along the axis yields
\be 
\begin{aligned}
\left(\int_{-\infty}^{-s_0} +\int_{s_0}^{\infty} \right)  \frac{{\rm d}s }{2\pi i} \frac{ \textrm{Disc}[{\cal M}(s)]}{s^3} 
&= -\int_{s_0}^{\infty}\frac{{\rm d}s }{2\pi i} \frac{ \textrm{Disc}[{\cal M}(-s)]}{s^3} +\int_{s_0}^{\infty} \frac{{\rm d}s }{2\pi i} \frac{ \textrm{Disc}[{\cal M}(s)]}{s^3}\\
&=
2\int_{s_0}^{\infty} \frac{{\rm d}s }{2\pi i} \frac{ \textrm{Disc}[{\cal M}(s)]}{s^3}.
\label{eq:crosssym}
\end{aligned}
\ee
Because the external states are identical, crossing symmetry implies that ${\cal M}(s+i\epsilon) = {\cal M}(-s-i\epsilon)$, so $\textrm{Disc}[{\cal M}(-s)] = - \textrm{Disc}[{\cal M}(s)]$.   Inserting the optical theorem, $\textrm{Im}[{\cal M}(s)] = s \sigma(s)$, the dispersion relation becomes\footnote{Note that in 3D, ${\cal M}(s)$ and $\sigma(s)$ have mass dimensions $+1$ and $-1$, respectively.}
\be 
16 a'  = \frac{2}{\pi} \int_{s_0}^{\infty}  {\rm d}s \frac{\sigma(s)}{s^2} \geq 0,
\ee
where $\sigma(s)$ is the total cross-section.  In the last step we have used the fact that the total cross-section is non-negative, implying that $a'\geq 0$.

The above arguments apply provided that high-energy scattering amplitudes comply with the optical theorem, the Froissart bound, and the standard analyticity properties of the S-matrix.   
The first and third conditions hold under the assumptions of unitarity and locality, respectively, while the second requires both.  In \Ref{Giddings}, it was noted that locality may break down when quantum
gravitational dynamics become important; in particular, black holes
may induce non-localities at super-Planckian energies, which violate the
Froissart bound and the polynomial boundedness of amplitudes \cite{Giddings}, albeit
in unphysical regions of complex momentum space \cite{Porto}. However, these caveats are immaterial because, as previously noted, black holes do not exist in asymptotically-flat 3D spacetime, so our arguments apply.  In 4D, the issue is more complex, but we postpone a dedicated discussion to \Sec{sec:Analyticity4}. 
 
\subsection{Unitarity (3D)}\label{sec:Unitarity3}

We now derive effective theory bounds by imposing unitarity on a general parameterization of the ultraviolet completion.  Our analysis follows the approach of \Ref{Dvali}.  As a consequence of the shift symmetry of $\phi$,  the leading coupling to high-energy degrees of freedom is uniquely
\be 
\chi_{\mu\nu} \partial^\mu \phi \partial^\nu \phi,
\label{eq:chicoupling}
\ee
where $\chi_{\mu\nu}$ is a field representing arbitrary ultraviolet dynamics.    
Integrating out these states generates the leading four-derivative operator, $(\partial_\mu \phi\partial^\mu \phi)^2$.   By neglecting higher-order interactions of $\phi$ with $\chi_{\mu\nu}$, we are implicitly assuming a perturbative ultraviolet completion.   Couplings of the form $ \chi_\mu\partial^\mu \phi $ are also allowed in principle but can be eliminated via the transverse condition  $\partial_\mu \chi^\mu =0$. Moreover, couplings of the form $ \partial_\mu \chi_\nu \partial^\mu \phi \partial^\nu \phi$ can be neglected because they produce subleading six-derivative operators.

We now decompose $\chi_{\mu\nu}$ into components,
\be 
\chi^{\vphantom{()}}_{\mu\nu} =  \chi^{(2)}_{\mu\nu} + \eta_{\mu\nu} \chi^{(0)} ,
\ee 
where $\chi^{(2)}_{\mu\nu}$ is by definition traceless.  In our conventions, all coupling constants have been absorbed into the overall normalization of the fields, so the leading interactions are of unit strength, $\chi^{(2)}_{\mu\nu} \partial^\mu \phi \partial^\nu \phi + \chi^{(0)} \partial_\mu \phi \partial^\mu \phi$.   Without loss of generality, the non-perturbative spectral representation of the $\chi_{\mu\nu}$ propagator in $D$ dimensions is given by
\be
\begin{aligned}
\langle  \chi^{(0)}(k) \chi^{(0)}(k') \rangle &=i \delta^D(k +k') \int_0^\infty {\rm d}\mu^2 \, \frac{ \rho^{(0)}(\mu^2) }{k^2- \mu^2+ i\epsilon}\\
\langle   \chi^{(2)}_{\mu\nu}(k)  \chi^{(2)}_{\rho\sigma}(k') \rangle &=i \delta^D(k+k')  \int_0^\infty {\rm d}\mu^2 \; \frac{ \rho^{(2)}(\mu^2) }{k^2- \mu^2+ i \epsilon} \Pi_{\mu\nu\rho\sigma},\label{eq:correlation3d2}
\end{aligned}
\ee
where $\rho^{(0)}$ and $\rho^{(2)}$ are spectral densities describing an arbitrary collection of single- or multi-particle intermediate states.  As usual, these expressions are obtained by inserting a complete set of states into the two-particle correlation function, implying positive definite spectral densities in the absence of tachyon or ghost instabilities.    Note also that since we are ultraviolet-completing a local operator, {\it i.e.}, one that is regular as $k\rightarrow 0$, the spectral density must vanish in the neighborhood of $\mu^2 =0$. 

As is well known, the spectral representation of a massive spin-2 state is strongly constrained by unitarity.   In $D$ dimensions, the absence of tachyons or ghosts implies that \cite{Hinterbichler}
\be 
 \Pi_{\mu\nu\rho\sigma}= \frac{1}{2}(\Pi_{\mu\sigma}\Pi_{\nu\rho}+\Pi_{\mu\rho}\Pi_{\nu\sigma}) - \frac{1}{D-1}\Pi_{\mu\nu}\Pi_{\rho\sigma},
 \label{eq:gravnum}
\ee
where for convenience we  have defined the projection operator,
\be 
\Pi_{\mu\nu} = \eta_{\mu\nu} -k_\mu k_\nu / \mu^2,
\label{eq:projection}
\ee  
such that $k^\mu \Pi_{\mu\nu} =0$ when $k^2 = \mu^2$ is on-shell.  Note that the transverse condition, $k^\mu \Pi_{\mu\nu\rho\sigma} =0$, applies on-shell so as to eliminate gauge degrees of freedom.  Not coincidentally, \Eq{eq:gravnum} is precisely the propagator numerator for a massive graviton. 

At low momentum transfer we integrate out $\chi_{\mu\nu}$, yielding the 3D effective operator,
\be 
  \chi_{\mu\nu}\partial^\mu \phi \partial^\nu \phi   \rightarrow  (\partial_\mu \phi \partial^\mu \phi )^2 \int_0^\infty {\rm d}\mu^2 \; \frac{\rho^{(0)}(\mu^2)/2 + \rho^{(2)} (\mu^2)/4}{\mu^2} .
\label{eq:IR3D}
\ee
\Eq{eq:IR3D} shows that the coefficient of $ (\partial_\mu \phi \partial^\mu \phi )^2$ is positive for any weakly-coupled ultraviolet completion consistent with a positive spectral density.   Thus, unitarity implies that $a' \geq 0$ for the effective Lagrangian defined in \Eq{eq:scalarL}.  Conversely, $a' <0$ signals an instability coming from a tachyon or ghost intermediate state.

The above arguments apply assuming a perturbative ultraviolet completion of the effective theory.  This allowed us to ignore operators involving ever higher powers of the field.
As discussed in \Sec{sec:Analyticity3}, while it may be problematic to extrapolate any argument to energies far above the Planck scale, this is not an issue in asymptotically-flat 3D spacetime, since black holes are not permitted.  

\subsection{Causality  (3D)}\label{sec:Causality3}

Let us now investigate the causal structure of the 3D photon-graviton effective theory.  We expand around non-trivial backgrounds for the photon and graviton,
\be 
\phi= \overline \phi + \varphi \qquad{\rm and}\qquad
g_{\mu\nu} = \overline g_{\mu\nu} + h_{\mu\nu}.
\label{eq:expand}
\ee 
Throughout, any barred variable represents a field or combination of fields evaluated on its background value.  Here $\varphi$ denotes photon fluctuations, while in 3D, the graviton is non-dynamical so $h_{\mu\nu} =0$.  To simplify our analysis we introduce vielbein coordinates defined by $\eta_{ab} = \overline e_a^\mu \overline e_b^\nu \overline g_{\mu\nu}$, where $\eta_{ab}= \textrm{diag}(+1,-1,-1)$ is the flat space metric.  We use Latin and Greek indices to denote vielbein and metric coordinates, respectively.  Importantly, the speed measured in the vielbein frame corresponds to the physical speed measured by an observer in the coordinates of the local Lorentz frame.
In terms of these coordinates, the equation of motion for $\varphi$ in a background is
\be
\tilde \eta^{ab} \partial_a\varphi \partial_b\varphi = 0 \label{eom1},
\ee
where $\tilde \eta_{ab}$ is defined as the effective metric in the vielbein frame, obtained from \Eq{eq:scalarL}, 
\be 
\tilde \eta_{ab} = \eta_{ab} + 16a'  \overline{\partial_a  \phi \partial_b \phi} .
\ee 
We study the geometric-optics limit in which $\varphi$ is a plane wave perturbation of four-momentum $k_a = (k_0, \vec k)$, with wavelength far shorter than the characteristic length scale of the spacetime curvature.  In this case, the dispersion relation for the photon is simply
\be 
\tilde \eta^{ab} k_a k_b =0.
 \ee 
 For now, let us focus on the photon speed in a local neighborhood; we will consider the global effects of gravity shortly. 
 
The local speed of photon fluctuations varies depending on the choice of background.   The simplest possibility is a constant electromagnetic field, represented by a constant condensate that breaks Lorentz invariance: $\overline{\partial_a  \phi}=w_a=(w_0, \vec w)$.  The effective metric is then $\tilde \eta_{ab} = \eta_{ab} + 16 a' w_a w_b$.
Expanding at leading order in the small parameter $a'$, we obtain the propagation speed of photons,
\be 
v = \frac{k_0}{|\vec k |} = 1- 8a'(w_0-\vec w \cdot \hat k)^2,
\ee 
defining $\hat k = \vec k / |\vec k|$. Superluminal photon propagation occurs when $a'<0$.  

Another interesting background is a thermal gas of photons, which we consider henceforth.  For a thermal system, background fields should be evaluated as stochastic expectation values, so in general $\overline{\partial_a \phi \partial_b \phi} \neq \overline{\partial_a \phi }  \cdot  \overline{\partial_b \phi }$.  In particular, for a photon gas the electromagnetic field has zero average background value, $\overline{\partial_a  \phi} = 0$, but non-zero variance, $ \overline{\partial_a \phi \partial_b  \phi} \neq 0$.  In 3D, the pressure $p$ and energy density $\rho$ satisfy an equation of state $p=\rho/2$, where $\rho = {\zeta(3)}{}T^3/\pi$ for a gas at temperature $T$ \cite{BB3D1,BB3D2}.
For a scalar field, the energy-momentum tensor is
\be 
T^{ab} =  \partial^a  \phi\partial^b \phi -\frac{1}{2} \eta^{ab}  \partial_c \phi\partial^c \phi,
\ee
the background expectation value of which is  $\overline{T}^{ab} = {\rm diag}(\rho,p,p)$ in a thermal gas.
From this we deduce that  $ \overline{\partial_c  \phi\partial^c \phi }= -2\overline{T}^{a}_{\;\;a} = -2(\rho-2p) = 0$, so
\be
\overline{ \partial_a  \phi \partial_b\phi} = (3\delta_a^0 \delta_b^0 - \eta_{ab})\frac{\zeta(3)}{2\pi}T^3.\label{eq:dphiab}
\ee
Putting everything together, we obtain the effective metric for photon propagation,
\be 
\tilde \eta_{ab} = \eta_{ab} +  (3\delta_a^0 \delta_b^0 - \eta_{ab})\frac{8\zeta(3)}{\pi}a'T^3.
\ee
The presence of Kronecker delta functions signals the fact that a thermal background breaks Lorentz invariance while preserving isotropy.   Expanding at leading order in $a'$, we find that the speed of signal propagation is
\be
v =  \frac{k_0}{|\vec k|} = 1 - \frac{12\zeta(3)}{\pi} a'T^3. 
\label{eq:speed1}
\ee
As before, superluminal propagation occurs when $a'<0$.

Traditionally, superluminal propagation is taken to be a definitive signal of an underlying pathology.  However, this diagnosis neglects an important distinction between superluminal propagation in all reference frames versus a preferred frame.  The present construction is of the latter type, which as discussed in \Ref{IRUV} introduces oddities in the definition of initial conditions, but is not, strictly speaking, inconsistent.  

To demonstrate a true breakdown of causality, we must construct a closed signal trajectory in spacetime, {\it i.e.}, a closed causal curve (CCC).   To begin, consider a thermal gas of photons localized to a finite bubble in spacetime.  The interior of the bubble is described by a zero-curvature, 3D Friedmann--Robertson--Walker (FRW) metric
\be
{\rm d}s^2 = a(t)^2 \eta_{\mu\nu} {\rm d}x^\mu {\rm d}x^\nu ,
\label{eq:FRW3}
\ee
written in a form that is manifestly conformally flat.  Inside the bubble,  photons deviate from the light-cone by an amount prescribed by the vielbein speed in  \Eq{eq:speed1}.  Meanwhile, the vacuum region exterior to the bubble is locally flat because the Weyl tensor vanishes identically in 3D.   Consequently, photons are exactly luminal outside the bubble.

What about the boundary of the bubble?  Since  the interior and exterior spacetimes are conformally flat, regularity of the spacetime across the boundary implies that, in the thin-shell limit, the boundary region itself is parametrically close to conformal flatness. Moreover, one can imagine a boundary formed from ``stiff" matter with $\rho=p$, for which the Cotton tensor vanishes \cite{Fluid3d}, thus ensuring conformal flatness exactly.\footnote{The signal itself can be transferred across the boundary either by another particle species that does not interact with the boundary material or by photons through a very small aperture in the circular shell.}  In any case, a bubble of thermal photons is well described by a metric that is globally conformally flat,
\be
{\rm d}s^2 = \Omega^2\eta_{\mu\nu}{\rm d}x^\mu {\rm d}x^\nu, 
\label{eq:confmetric}
\ee
where $\Omega = 1$ in the exterior and $\Omega = a(t)$ in the interior.  A feature of conformal flatness is that the speeds of signals as measured in vielbein coordinates and metric coordinates are the same. That is, light signals move at speed $v={\rm d}x/{\rm d}t$, where $v$ is given by Eq. \eqref{eq:speed1}.\footnote{This is in marked contrast to the gravitational redshift of signals in spacetimes that are not conformally flat, such as a signal propagating radially away from a black hole.}    In the end, this implies that engineering a CCC in a conformally-flat spacetime reduces to a special relativistic problem.  As is well known, however, a CCC in special relativity requires two frames in relative motion, while the above construction picks out a single preferred frame.   
To build a CCC we must instead consider two bubbles of thermal photons, both at temperature $T$ and in relative motion.  The associated background is described by \Eq{eq:confmetric}, only with a more complicated form for the conformal factor.

\begin{figure}[t] 
\centerline{\includegraphics[width=.55\columnwidth]{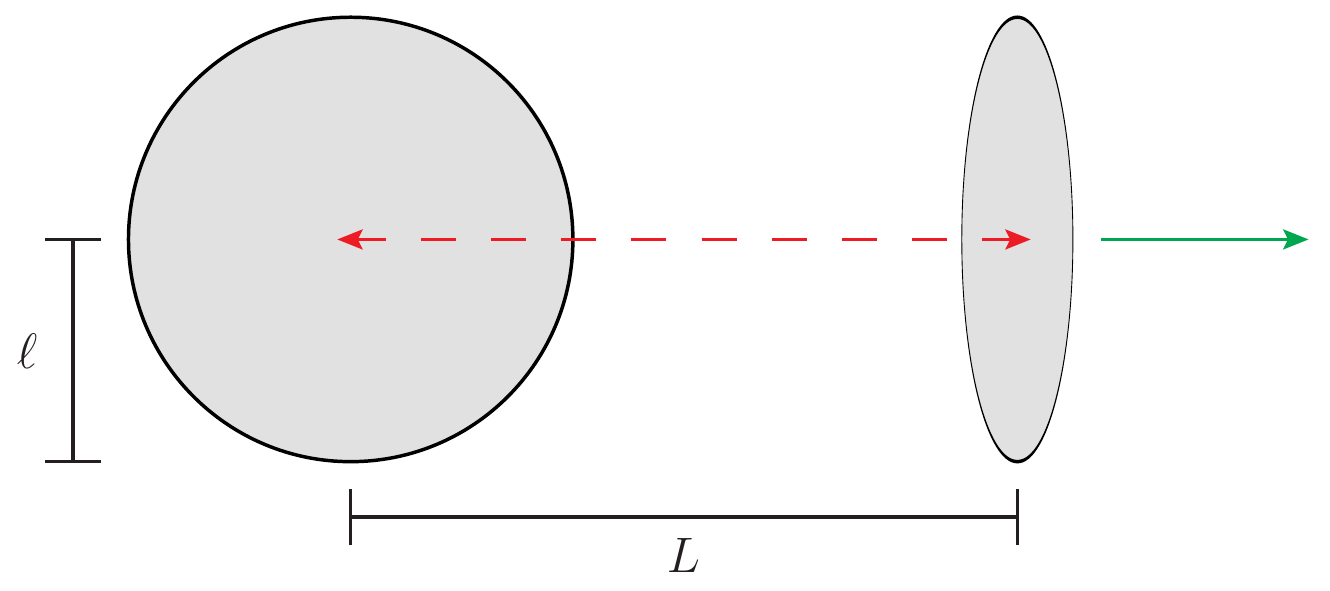}}
\caption{Setup for the construction of a CCC in 3D, using superluminal photons in a theory that violates \Eq{eq:consistency3d}. The construction, illustrated here in a constant-time slice of the two spatial dimensions, consists of two circular bubbles of thermal radiation, each of radius $\ell$ and separation $L\gg\ell$, with relative speed $u<1$ (green arrow). Signals (red dashed arrows) sent back and forth would be superluminal within the bubbles, creating a CCC for large $u$.}\label{fig:bubbles}
\end{figure}

Now consider the setup illustrated in \Fig{fig:bubbles}: two bubbles of equal radii $\ell$ separated by a distance $L$ and moving in opposite directions at zero impact parameter and constant speed $u$.   Light signals sent between observers at the center of each bubble will have an average speed 
\be
v_{\rm avg} = 1-\epsilon \label{eq:speedCCC},
\ee 
as measured in their respective frames.  Here, corrections to the speed of light are controlled by a small parameter, $\epsilon \sim a'T^3\ell/L$.   Note that the Friedmann equations imply that the interior of each bubble will evolve on a timescale $\sim \rho^{-1/2}$ in natural units.  However, these effects can be neglected by choosing  $L^2  T^3\ll 1$, which is always possible for sufficiently small $T$.  Consequently, we  can always treat the temperature as roughly constant over the entire signal trip.

For $a'<0$, it is then straightforward to construct a CCC.  Explicitly, each observer can send a signal that in the reference frame of the other observer propagates at an average superluminal speed defined by \Eq{eq:speedCCC}.  By transmitting a signal from one bubble to the other and then back, it is possible to form a CCC.  This is analogous to the so-called ``tachyonic antitelephone'' from special relativity \cite{Tolman,antitelephone,Bohm}.  Likewise, causality violation will occur here provided the relative speed of the two observers ({\it i.e.}, the relative speed of the bubbles) satisfies
\be 
u>\frac{2v_{\rm avg}}{1+v_{\rm avg}^2} \simeq 1-\frac{1}{2}\epsilon^2.\label{eq:compatibilityCCC}
\ee
A diagram of this CCC is nicely depicted in Fig.~2 of \Ref{IRUV}, albeit in a slightly different context (Lorentz-violating condensate bubbles passing with finite impact parameter) and without including the effects of gravity. Forbidding the existence of causality violation from a CCC thus requires $a'\geq0$.

The above arguments apply provided there is no subtlety in constructing this particular background of thermal photons.  Na\"ively, one may worry about exceeding the limits of the photon-graviton effective theory due to the relative boost between the bubbles of gas. However, this is not an issue because the bubbles need not overlap and hence do not back-react.   While arbitrary configurations of moving masses in 3D sometimes entail topological subtleties \cite{3dtimemachine1,3dtimemachine2,DeserCTC2}, our CCC construction does not rely on them for causality violation.  A detailed study of these issues goes beyond the scope of the current paper.

\section{Four Dimensions}
\label{sec:4d}
\subsection{Setup and Bounds (4D)}

In this section, we derive bounds on the photon-graviton effective action in 4D.  As in \Sec{sec:eff3d}, we rewrite the spacetime curvature in terms of the electromagnetic field strength.  To start, we eliminate $R_{\mu\nu\rho\sigma}R^{\mu\nu\rho\sigma}$ from \Eq{eq:EH} in favor of the 4D Gauss--Bonnet term,
\be 
R^2 - 4 R_{\mu\nu} R^{\mu\nu} + R_{\mu\nu\rho\sigma}R^{\mu\nu\rho\sigma},
\ee
which is in turn a total derivative. We also use the definition of the Weyl tensor in \Eq{eq:Weyldef} to rewrite the operator $F_{\mu\nu}F_{\rho\sigma}R^{\mu\nu\rho\sigma}$ in terms of $F_{\mu\nu}F_{\rho\sigma}C^{\mu\nu\rho\sigma}$, $F_{\mu\rho}F_\nu^{\;\;\rho}R^{\mu\nu}$, and $F_{\mu\nu}F^{\mu\nu}R$. Next, we  substitute the energy-momentum tensor \eqref{eq:T} in for the Ricci scalar and Ricci tensor in the higher-dimension operators using the tree-level Einstein field equations \eqref{eq:Einstein}, which at the present order in couplings is again equivalent to a field redefinition.  With the useful identity in 4D,
\be
2(F_{\mu\nu}F^{\mu\nu})^2 + (F_{\mu\nu}\tilde{F}^{\mu\nu})^2 = 4F_{\mu\rho} F_{\nu}^{\;\;\rho}F^{\mu}_{\;\;\sigma} F^{\nu \sigma},\label{eq:usefulidentity}
\ee
we obtain our final form for the effective Lagrangian, 
\be 
 \mathcal{L} = -\frac{1}{4} F_{\mu\nu} F^{\mu\nu} - \frac{1}{4} R 
 + a_1' (F_{\mu\nu} F^{\mu\nu})^2 +  a_2' (F_{\mu\nu} \tilde F^{\mu\nu})^2 + b_3 F_{\mu\nu} F_{\rho\sigma} C^{\mu\nu\rho\sigma} ,
\label{eq:EH2}
\ee
where we have defined new higher-dimension operator coefficients, 
\be 
a_1' = a_1 - b_2/2  -b_3 +c_2+ 4c_3  \qquad {\rm and} \qquad 
a_2' = a_2 - b_2/2  -b_3 +c_2+ 4c_3.
\label{eq:a12p}
\ee
In \Eq{eq:EH2}, all explicit curvature dependence has been removed except for the Weyl tensor, which in 4D is non-trivial.  
In the classical theory, the Weyl tensor represents the component of the gravitational field that propagates freely in the absence of sources and thus decouples from matter at leading order in Einstein's equations.  Later, we will see how this is  manifested in the forward scattering amplitudes, which at leading order are explicitly dependent on $a_i'$ but not $b_3$.  
 
Constraining the parameters in \Eq{eq:EH2} using analyticity, unitarity, and causality is substantially more difficult in 4D due to dynamical gravity.  We will elaborate on these arguments later on, but let us briefly collect our final results here.  We derive bounds coming from unitarity:
\be 
a_1'  \geq  0  \qquad{\rm and}\qquad a_2'\geq 0,
\label{eq:unitarity4}
\ee
while the absence of superluminality in certain backgrounds implies that
\be 
a_1' +a_2'\geq 0.
\label{eq:superluminality4}
\ee
Interestingly, if one blithely applies analyticity arguments to the higher-dimension operator contributions, one also obtains \Eq{eq:superluminality4}. 
Just as in \Sec{sec:eff3d}, it is convenient to expand $a_i'$ in terms of their contributions from electromagnetic and gravitational interactions:
\be 
a_i' = \alpha_i z^4 + \beta _i z^2 + \gamma_i  ,
\ee
where $\alpha_i$, $\beta_i$, and $\gamma_i$ are generated by diagrams like the ones shown in \Fig{fig:abc}.
Contributions coming from integrating out a charged fermion \cite{D&H} or charged scalar \cite{Dunne,Bastianelli1,Bastianelli2} are
\be
\begin{aligned}
(a_1,a_2) &= \left(\frac{q^4}{1440 \pi^2 m^4 } , \;
 \frac{7q^4}{5760 \pi^2 m^4 } \right) && {\rm [fermion]}\\ 
(b_1,b_2,b_3) &= \left(-\frac{q^2}{576 \pi^2 m^2 },  \;\frac{13q^2}{1440 \pi^2 m^2 }, \;
 -\frac{q^2}{1440 \pi^2 m^2 } \right) && {\rm [fermion]} \\
(a_1,a_2) &= \left(\frac{7q^4}{23040 \pi^2 m^4 } , \;
 \frac{q^4}{23040 \pi^2 m^4 } \right) && {\rm [scalar]} \\
(b_1,b_2,b_3) &= \left(\frac{q^2}{1152 \pi^2 m^2 },  \;\frac{q^2}{1440 \pi^2 m^2 }, \;
 \frac{q^2}{2880 \pi^2 m^2 } \right) && {\rm [scalar]},
\end{aligned} 
 \label{eq:abc4}
\ee
where for the scalar we have assumed minimal coupling to gravity.  Given these coefficients, the unitarity bounds in \Eq{eq:unitarity4} imply that 
\be  
\begin{aligned}
z^2\left(z^2-11/2 \right)& \geq - \gamma_1 \times 1440 \pi^2 &&  {\rm [fermion]} \\
z^2\left(z^2-22/7 \right) & \geq - \gamma_2 \times 5760 \pi^2 /7  && {\rm [fermion]} \\
z^2\left(z^2-16/7 \right)& \geq - \gamma_1 \times 23040 \pi^2 /7 &&  {\rm [scalar]} \\
z^2\left(z^2-16 \right) & \geq - \gamma_2 \times 23040 \pi^2  && {\rm [scalar]} ,\label{eq:UnitarityBound4D}
\end{aligned}
\ee
while the bounds from analyticity and superluminality in \Eq{eq:superluminality4} are
\be  
\begin{aligned}
z^2\left(z^2-4\right)& \geq -(\gamma_1 + \gamma_2)\times 5760\pi^2 / 11 &&  {\rm [fermion]} \\
z^2\left(z^2-4\right) & \geq -(\gamma_1 + \gamma_2)\times 2880\pi^2  && {\rm [scalar]}.\label{eq:CausalityBound4D}
\end{aligned}
\ee
Curiously, for small values of $\gamma_i$, both fermions and scalars in 4D are subject to the same bound:
\be
\begin{aligned}
z\geq 2.
\end{aligned}
\ee

\begin{figure}[t] 
\centerline{\includegraphics[width=1.\columnwidth]{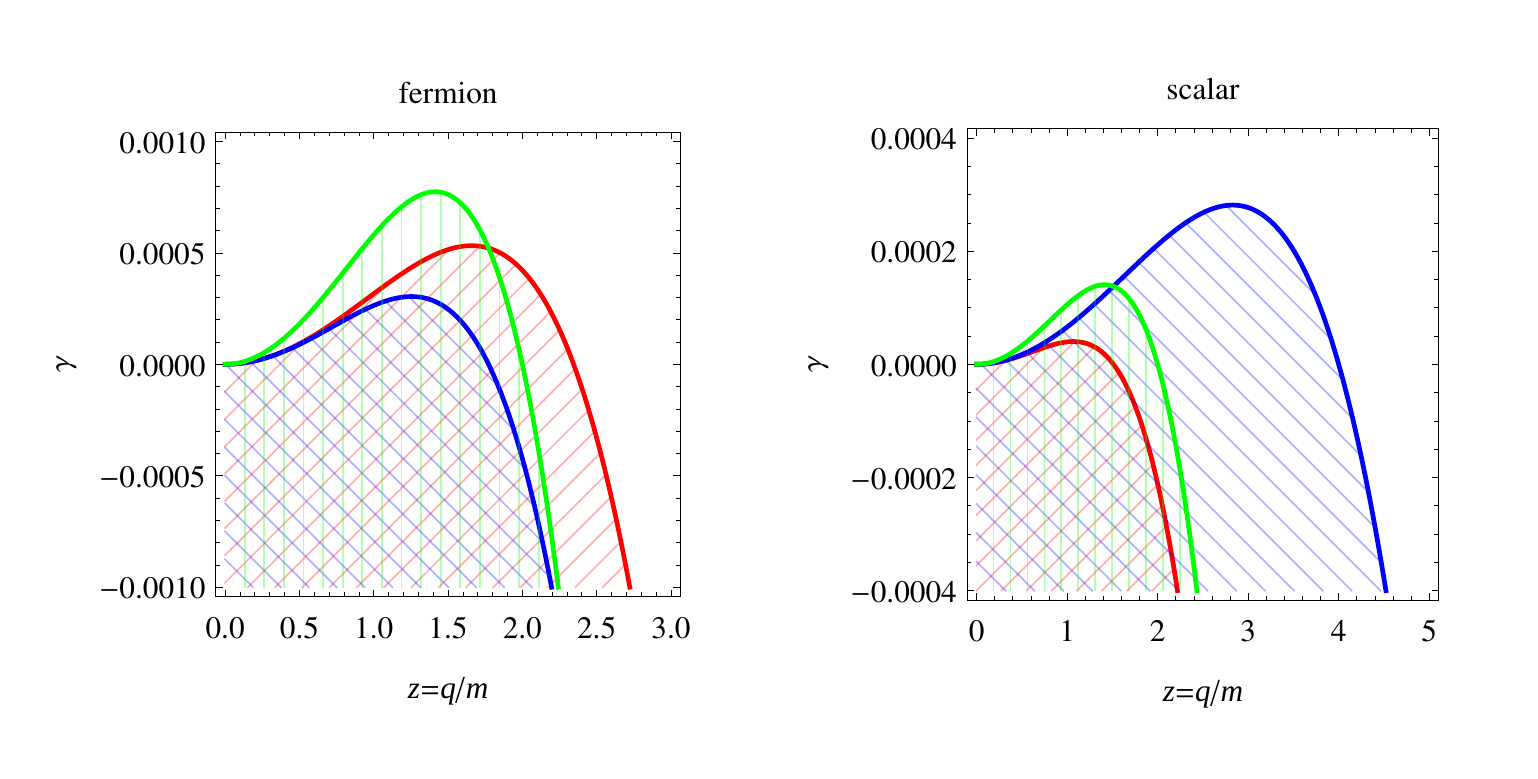}}\vspace*{-0.75cm}
\caption{Bounds on the 4D photon-graviton effective theory derived from integrating out a fermion (left) or scalar (right) and expressed in terms of the contributions coming from electromagnetism (parameterized by $z=q/m$) and pure gravity (parameterized by $\gamma$).   The cross-hatched regions are forbidden by arguments from unitarity, which apply to $\gamma =\gamma_1$ ({\color{red} red $\boxslash$}) and $\gamma=\gamma_2$ ({\color{blue} blue $\boxbslash$}), and arguments from analyticity and superluminality, which both apply to $\gamma= \gamma_1+\gamma_2$ ({\color{green} green $\boxbar$}).  The WGC forbids $z<1$, which overlaps with much of the region also forbidden by infrared consistency.
}\label{fig:bound4d}
\end{figure}

All of our 4D constraints are summarized in \Fig{fig:bound4d}. As in \Sec{sec:eff3d}, the coefficients $\gamma_i$ parameterize all corrections coming from purely gravitational interactions.  In 4D, this includes the contribution from $c_2 +4c_3$ in \Eq{eq:a12p}, which runs logarithmically due to graviton loops \cite{Stan} and is thus controlled by an ultraviolet-sensitive boundary condition. 
As in 3D, for sufficiently large values of $\gamma_i$ these bounds are automatically satisfied. 
Alternatively, we can consider the case where the purely Planck-suppressed corrections are negligible, in which case $\gamma_i$ is small and  our infrared consistency conditions bound $z$ strictly from below.

It is reasonable to assume that a theory that satisfies our consistency conditions, \Eqs{eq:UnitarityBound4D}{eq:CausalityBound4D}, at a given energy scale will continue to do so deeper into the infrared.  Interestingly, this implies that $\gamma_i$ should not decrease in the infrared, {\it i.e.}, the sign of the beta function for $\gamma_i$ should be negative on general grounds.  This is confirmed by explicit computation of the one-loop divergences in the photon-graviton effective theory \cite{Stan}.

\subsection{Analyticity (4D)}\label{sec:Analyticity4}
Let us endeavor to apply the analyticity argument of \Sec{sec:Analyticity3} to light-by-light scattering in 4D. Using \Eq{eq:EH2}, we read off the Feynman rules for the photon-graviton theory: there is the usual photon-photon-graviton vertex from the Einstein--Maxwell terms, a higher-order photon-photon-graviton vertex from the $b_3$ term, and new quartic photon vertices from the $a_1'$ and $a_2'$ terms. Putting these together, we find that the tree-level helicity amplitudes are
\be
\begin{aligned}
\mathcal{M}({1^+2^+3^+4^+} )=\mathcal{M}({1^-2^-3^-4^-} ) &= 8(a_1'-a_2')(s^2+t^2+u^2) \\
\mathcal{M}({1^+2^+3^-4^-}) = \mathcal{M}({1^-2^-3^+4^+}) &= \frac{2s^4}{stu}+8(a_1' + a_2')s^2 \\
\mathcal{M}({1^+2^-3^+4^-}) = \mathcal{M}({1^-2^+3^-4^+})  &= \frac{2t^4}{stu}+8(a_1' + a_2')t^2 \\
\mathcal{M}({1^+2^-3^-4^+}) = \mathcal{M}({1^-2^+3^+4^-})  &= \frac{2u^4}{stu}+8(a_1' + a_2')u^2 ,
\end{aligned}\label{eq:4damplitude}
\ee
where all remaining amplitudes are given by permutations of
\be
\begin{aligned}
\mathcal{M}({1^+2^+3^+4^-}) &= \mathcal{M}({1^-2^-3^-4^+}) =  2b_3(s^2+t^2+u^2).
\end{aligned}\label{eq:4damplitudeb}
\ee
Here, we have used a helicity basis defined with all momenta incoming, so the second and fourth lines of \Eq{eq:4damplitude} correspond to forward amplitudes.   Notably, at leading order in the higher-dimension operator coefficients, all forward amplitudes depend explicitly on $a_i'$ but not $b_3$, which controls the irreducible interactions between the electromagnetic field strength and the Weyl tensor. 
This is quite reasonable on physical grounds because, in the classical limit, the Weyl tensor does not have a minimal coupling to the energy-momentum tensor.  Quantum mechanically, this is manifested as the fact that the Weyl tensor mediates forward light-by-light scattering only at second order in $b_3$, {\it i.e.}, coming from two insertions of the higher-dimension operator.

Critically, the scattering amplitudes have terms that are singular in the $s$-, $t$-, and $u$-channels due to leading-order graviton exchange.  In the forward limit, the $t$-channel diagrams scale as $ \sim s^2/t$ and formally diverge at forward scattering.  In this limit,
the partial wave expansion does not converge, the Froissart bound is invalid, and the dispersion relation reasoning from \Sec{sec:Analyticity3} does not apply.  Hence, dynamical gravity creates a considerable obstacle to any argument from analyticity \cite{IRUV}.  

There is no immediate justification for simply dropping these singular contributions.  Nevertheless, it is interesting to compute the bound that would arise from applying the analyticity argument of \Sec{sec:Analyticity3} to the non-singular contributions coming from higher-dimension operators.  Notably, a crucial ingredient of the analyticity argument is the requirement that contributions to the scattering amplitude from ultraviolet dynamics be even in $s$.  As a result, contributions to the dispersion relation from negative $s$ can be directly related to the cross-section at positive $s$.  In 3D, this was automatically satisfied because the scattering amplitude was a crossing-symmetric function characterizing indistinguishable scalars.  In contrast, the 4D scattering amplitudes describe photons with distinguishable helicity labels.  To form an object suitable for analyticity bounds, we consider the sum of all forward amplitudes, $\mathcal{M}_{\rm sym}$, which is by construction symmetric under the exchange of $1\leftrightarrow 3$ and $2\leftrightarrow 4$:
\be
\begin{aligned}
\mathcal{M}_{\rm sym} &=\mathcal{M}(1^+ 2^+ 3^- 4^-)  +\mathcal{M}(1^- 2^- 3^+ 4^+)  +\mathcal{M}(1^+ 2^- 3^- 4^+)  +\mathcal{M}(1^- 2^+ 3^+ 4^-)  \\
&= \dfrac{4(s^4+u^4)}{stu}+16(a_1' + a_2')(s^2+u^2) \\
& \overset{t\rightarrow 0}{=}  -\dfrac{8s^2}{t}-8s+32(a_1' + a_2')s^2 + \mathcal{O}(t) .
\end{aligned}
\ee
The first two terms of the last line come from single graviton exchange due to the Einstein--Maxwell photon-photon-graviton vertex. If we drop this contribution, then the dispersion relation argument of \Sec{sec:Analyticity3} implies that the coefficient of $s^2$ in $\mathcal{M}_{\rm sym}$ is non-negative, so
\bea
a_1' +a_2' \geq 0.
\label{eq:4dan}
\eea
Because the $t$-channel graviton singularity remains a critical obstruction to this argument, the inequality in \Eq{eq:4dan} should not yet be considered a rigorous bound.   Nevertheless, it has been noted that singular contributions can be consistently subtracted  from a dispersion relation \cite{IRUV,Bellazzini}, provided the theory has a weak coupling parameter that can discriminate between the  contribution from leading-order exchange of massless particles and that from higher-dimension operators.
For the photon-graviton effective action, the natural choice for a weak coupling parameter is the gravitational constant, $G$. However, by sending $G \rightarrow 0$, we also eliminate the very higher-dimension, gravitationally-induced interactions that we seek to bound.  Thus, we have not identified such a weak coupling parameter here, though it may be possible.  More generally, it may be feasible to extract rigorous effective theory bounds from theories with $t$-channel singularities, but we leave this formidable task for future work.

As discussed in \Sec{sec:Analyticity3}, the analyticity argument involves additional subtleties related to taking a contour in the complex $s$ plane to super-Planckian scales, which {\it a priori} could involve issues with black hole formation and associated non-localities. However, as $t\rightarrow 0$, the impact parameter exceeds the Schwarzschild radius for the scattering particles, implying no black hole production in the forward limit \cite{Porto,Dimopoulos}. Pathologies associated with non-perturbative gravitational interactions are thus avoided.  In any case, the same assumptions used in our analyticity bounds, which were mentioned in \Sec{sec:Analyticity3}, have been used previously to constrain string theories from low-energy scattering \cite{IRUV,Rothstein}. In general, this is justified because string amplitudes are analytic and highly convergent at large $s$ \cite{Pomeron, BCFWString}.

\subsection{Unitarity (4D)}\label{sec:Unitarity4} 
 
Next, let us apply the unitarity argument of \Sec{sec:Unitarity3} to 4D.  In principle, one can define general spectral representations parameterizing the ultraviolet-completing dynamics of $(F_{\mu\nu}F^{\mu\nu})^2$, $(F_{\mu\nu}\tilde F^{\mu\nu})^2$, and $F_{\mu\nu} F_{\rho\sigma} C^{\mu\nu\rho\sigma}$.  The only substantive difference from the 3D case is the third operator, which depends on the spacetime curvature in a way that cannot simply be eliminated using Einstein's equations.  In what follows, we will be interested in bounding the coefficients of the first and second operators.

At leading order, the photon couples to the ultraviolet states according to
\be 
F^{\mu\nu} F^{\rho \sigma} \chi_{\mu\nu\rho\sigma}  \qquad{\rm and}\qquad F^{\mu\nu} \tilde F^{\rho\sigma} \psi_{\mu\nu\rho\sigma},\label{eq:psichi}
\ee
where $\chi_{\mu\nu\rho\sigma}$ and $\psi_{\mu\nu\rho\sigma}$ are parity-even and -odd fields that couple to the photon.   Note that these fields have the skew and interchange index symmetries of the Riemann tensor: $\chi_{\mu\nu\rho\sigma} = -\chi_{\nu\mu\rho\sigma} = -\chi_{\mu\nu\sigma\rho}$ and $\chi_{\mu\nu\rho\sigma} = \chi_{\rho\sigma\mu\nu}$ and similarly for $\psi_{\mu\nu\rho\sigma}$.   As in \Sec{sec:Unitarity3}, $\chi_{\mu\nu\rho\sigma}$ and $\psi_{\mu\nu\rho\sigma}$ parameterize an arbitrary set of intermediate single- or multi-particle states, so our unitarity argument remains quite general.  

While there can also exist couplings of the form $\chi_{\mu\nu} F^{\mu\nu}$, they can be eliminated by the transverse condition, $\partial_\mu \chi^{\mu\nu}=0$. Likewise, couplings of the form $\partial_\mu \partial_\nu \partial_\rho \chi_\sigma F^{\mu\nu}F^{\rho\sigma} $ and $\partial_\mu \chi_{\nu\rho\sigma} F^{\mu\nu}F^{\rho\sigma}$ need not be considered because they yield operators that are of higher order in the derivative expansion.  
Because the photon-graviton effective action includes the operator $F_{\mu\nu} F_{\rho\sigma} C^{\mu\nu\rho\sigma}$, it is also possible, in principle, that  $\chi_{\mu\nu\rho\sigma}$ couples directly to $C_{\mu\nu\rho\sigma}$.   However, as shown in \Sec{sec:Analyticity4}, interactions mediated through the Weyl tensor do not affect the low-energy forward scattering amplitudes at leading order in the higher-dimension operator coefficients.  Hence, at this order, any coupling between $\chi_{\mu\nu\rho\sigma}$ and $C_{\mu\nu\rho\sigma}$ cannot contribute to the coefficients of $(F_{\mu\nu}F^{\mu\nu})^2$ and $(F_{\mu\nu}\tilde F^{\mu\nu})^2$ and can be neglected.

As before, we expand $\chi_{\mu\nu\rho\sigma}$ into its components,
\be 
\chi^{\vphantom{()}}_{\mu\nu\rho\sigma} =  \chi^{(4)}_{\mu\nu\rho\sigma} + \frac{1}{4} (\eta^{\vphantom{()}}_{\mu[\rho} \chi^{(2)}_{\sigma]\nu}-\eta^{\vphantom{()}}_{\nu[\rho}\chi^{(2)}_{\sigma]\mu}) +\frac{1}{2} \chi^{(0)} \eta^{\vphantom{()}}_{\mu[\rho}\eta^{\vphantom{()}}_{\sigma]\nu},\label{eq:chi}
\ee
and similarly for  $\psi_{\mu\nu\rho\sigma}$, where $\chi^{(2)}_{\mu\nu}$ and $\chi^{(4)}_{\mu\nu\rho\sigma}$ are by definition traceless.  Also as in \Sec{sec:Unitarity3}, we choose a normalization in which all coupling constants are absorbed into the fields and the photon interacts via $ \chi^{(4)}_{\mu\nu\rho\sigma} F^{\mu\nu} F^{\rho\sigma} +\chi^{(2)}_{\mu\nu} F^{\mu}_{\;\; \rho} F^{\nu \rho} +\chi^{(0)} F_{\mu\nu} F^{\mu\nu}$.

The spectral decompositions for $\chi^{(0)}$ and $\chi^{(2)}_{\mu\nu}$ are the same as in \Eq{eq:correlation3d2}, while for $\chi^{(4)}_{\mu\nu\rho\sigma}$,
\be 
\langle  \chi^{(4)}_{\mu\nu\rho\sigma}(k) \chi^{(4)}_{\alpha\beta\gamma\delta}(k') \rangle = i \delta^D(k+k') \int_0^\infty {\rm d}\mu^2 \; \frac{ \rho^{(4)}(\mu^2) }{k^2- \mu^2 +i \epsilon} \Pi_{\mu\nu\rho\sigma\alpha\beta\gamma\delta},
\label{eq:Pi4}
\ee
where $\rho^{(4)}$ is the spectral function for the four-index state.  {\it A priori}, the tensor numerator $\Pi_{\mu\nu\rho\sigma\alpha\beta\gamma\delta}$ consists of arbitrary combinations of $\eta_{\mu\nu}$ and $k_\mu$; however, it is actually very constrained.   By construction, $\Pi_{\mu\nu\rho\sigma\alpha\beta\gamma\delta}$ is traceless with index (anti-)symmetry properties consistent with those of $\chi^{(4)}_{\mu\nu\rho\sigma}$.  In addition, just as for the spin-2 case, there are general arguments that fix the form of $\Pi_{\mu\nu\rho\sigma\alpha\beta\gamma\delta}$.    As discussed in \Refs{Weinberg1}{Weinberg3}, the tensor numerators of higher-spin propagators are functions of the projection operator $\Pi_{\mu\nu}$ defined in \Eq{eq:projection}.  This ensures that the transverse condition $k^\mu  \Pi_{\mu\nu\rho\sigma\alpha\beta\gamma\delta} = 0$ applies on-shell.  This is analogous to the usual transverse conditions required for theories of massive higher-spin fields.  We have checked that the only projection operator that satisfies the requisite trace, index symmetry, and transverse conditions can indeed be written in terms of combinations of $\Pi_{\mu\nu}$ and is moreover comprised of two such linearly independent tensor structures, shown in \Eqs{eq:bigPi1}{eq:bigPi2} of \App{appendix}.   Last of all, unitarity implies that \cite{Schwartz}
\be 
\Pi_{\mu\nu\rho\sigma\alpha\beta\gamma\delta} = \sum_i  \varepsilon_{i\mu\nu\rho\sigma}\varepsilon^*_{i\alpha \beta\gamma \delta},
\ee
so the tensor numerator is equal to the sum over polarization tensors labeled by $i$, with normalization $ \varepsilon_{i\mu\nu\rho\sigma}\varepsilon_{j}^{*\mu\nu\rho\sigma} = \delta_{ij}$.  However, the tensor numerator shown in \Eqs{eq:bigPi1}{eq:bigPi2} of \App{appendix} identically satisfies  $\Pi_{\mu\nu\rho\sigma}^{\;\;\;\;\;\;\;\;\; \mu\nu\rho\sigma} =0$, indicating that $\chi_{\mu\nu\rho\sigma}^{(4)}$ carries states of negative norm.  Thus, we conclude that $\chi^{(4)}_{\mu\nu\rho\sigma}$ is unphysical and should be eliminated altogether.

Nonetheless, $\chi^{(0)}$ and $\chi^{(2)}_{\mu\nu}$ are still propagating and unitarity dictates that their spectral functions $\rho^{(0)}$ and $\rho^{(2)}$ be positive.  At low energies, integrating them out yields
\be 
\chi_{\mu\nu\rho\sigma}F^{\mu\nu}F^{\rho\sigma}  \rightarrow ( F_{\mu\nu}F^{\mu\nu} )^2 \int_0^\infty {\rm d}\mu^2 \; \frac{\rho^{(0)}/2 + \rho^{(2)} /12}{\mu^2} + (F_{\mu\nu}\tilde F^{\mu\nu})^2 \int_0^\infty {\rm d}\mu^2 \; \frac{\rho^{(2)}/8}{\mu^2}.
\label{eq:IR}
\ee
 Thus, the contributions to $(F_{\mu\nu}F^{\mu\nu})^2$ and $(F_{\mu\nu}\tilde F^{\mu\nu} )^2 $ are both positive.
 
An analogous argument applies to the parity-odd field, $\psi_{\mu\nu\rho\sigma}$.  To see this, we define 
\be 
\psi_{\mu\nu\rho\sigma} F^{\mu\nu}\tilde{F}^{\rho\sigma}
  = \tilde{\chi}_{\mu\nu\rho\sigma} F^{\mu\nu}F^{\rho\sigma},
\ee 
where $\tilde{\chi}_{\mu\nu\rho\sigma}=\epsilon_{\;\;\;\,\rho\sigma}^{\alpha\beta}\psi_{\mu\nu\alpha\beta}/2$ is a parity-even field with the exact same symmetries as $\chi_{\mu\nu\rho\sigma}$. Running through the same logic as above implies that integrating out $\psi_{\mu\nu\rho\sigma}$ induces positive coefficients for $(F_{\mu\nu}F^{\mu\nu})^2$ and $(F_{\mu\nu}\tilde F^{\mu\nu} )^2 $.  Putting it all together, we find that unitarity implies 
\be 
a_1' \geq 0 \qquad{\rm and}\qquad a_2' \geq 0
\ee
for a weakly-coupled ultraviolet completion free of ghosts or tachyons.  

\subsection{Causality (4D)}

We now turn to the problem of calculating the speed of photon propagation in a non-trivial 4D background.   As before, we implement perturbation theory around a background electromagnetic and gravitational field,
\be 
A_\mu = \overline A_\mu + {a}_\mu , \qquad
g_{\mu\nu} = \overline g_{\mu\nu} + h_{\mu\nu},
\ee 
where the graviton is fully dynamical in 4D.  Similarly, the electromagnetic field strength can be expanded as $F_{\mu\nu} = \overline{F}_{\mu\nu} + f_{\mu\nu}$, with $f_{\mu\nu} =\overline{\nabla}_\mu a_\nu- \overline{\nabla}_\nu a_\mu = \partial_\mu a_\nu- \partial_\nu a_\mu$, where the final equality follows from the cancellation of the connection coefficients in the covariant derivatives.

Expanding perturbatively in the photon is straightforward for $(F_{\mu\nu}F^{\mu\nu})^2$ and $(F_{\mu\nu} \tilde F^{\mu\nu})^2$, but a slight subtlety arises for $F_{\mu\nu}F_{\rho\sigma}C^{\mu\nu\rho\sigma}$. 
In particular, this operator carries dependence on graviton fluctuations, which na\"ively can be eliminated in favor of the photon using the linearized Einstein field equations.  However, as discussed in \Secs{sec:Analyticity4}{sec:Unitarity4}, this does not actually happen because the Weyl tensor does not couple minimally to the energy-momentum tensor.  Thus, the graviton dependence in $F_{\mu\nu}F_{\rho\sigma}C^{\mu\nu\rho\sigma}$ can be dropped, although this operator still contributes to the photon dispersion relation through the Weyl tensor background value, $ \overline{C}_{\mu\nu\rho\sigma}$. This is nicely consistent with the analyticity arguments of \Sec{sec:Analyticity4} because of the close relationship between light-by-light scattering and the propagation of photons in a fixed electromagnetic background \cite{IRUV}.

Let us consider a photon fluctuation described by a plane wave with circular polarization $\varepsilon_a$ and momentum $k_a$.   Throughout, we work in Lorenz gauge, $k_a\varepsilon^a=0$. As before, we go to a geometric-optics limit in which the wavelength of the photon is far shorter than the typical scale of spacetime curvature \cite{MTW}.  In this regime, the dispersion relation is
\be 
\tilde \eta^{ab} k_a k_b = 0,
\ee 
where at leading order in the couplings $a'_{i}$ and $b_3$ the effective metric is 
\be 
\tilde \eta_{ab} = \eta_{ab} + 32 \left(a_1' \overline{F_{ac}F_{bd}}+a_2'  \overline{\tilde F_{ac} \tilde F_{bd}}\right) \varepsilon^{c*} \varepsilon^d + 8b_3 \overline{C}_{acbd}\varepsilon^{c*} \varepsilon^d.\label{eq:effmet4Weyl}
\ee 
Since the speed of propagation depends on the photon polarization, non-trivial electromagnetic fields induce birefringence.

 In analogy with \Sec{sec:Causality3}, it is natural to consider a constant electromagnetic background, $\overline F_{ab} \neq 0$, defined in vielbein coordinates.   However, an additional complication arises due to dynamical gravity: a non-trivial electromagnetic background induces photon-graviton mixing of the form $\overline F_{a}^{\;\;c} f_{bc} h^{ab}$.   This effect has been neglected in the literature on higher-order corrections to the photon dispersion relation \cite{Shore, D&H}, most likely because it is Planck-suppressed.  However, these corrections can easily dominate over contributions from higher-dimension operators in the photon-graviton effective action.  For example, in the range where the WGC is marginally satisfied, $m/q$ is of order the Planck scale and the effects of photon-graviton mixing will dwarf those of the higher-dimension operators.

To sidestep the issue of photon-graviton mixing, we focus on a background of thermal photons at temperature $T$.  Since the background field values are thermally averaged, $\overline{F_{ab} F_{cd}}\neq \overline{F}_{ab} \cdot \overline{F}_{cd}$.
In particular, for a photon gas, the electromagnetic field has zero average value, $\overline F_{ab} =0$, but non-zero variance, $\overline{ F_{ab}  F_{cd}}\neq 0$.  Photon-graviton mixing is identically zero because it scales as a single power of $\overline{F}_{ab}$.  Strictly speaking, this applies to quanta at wavelengths longer than $\sim 1/T$, so the effects of the background photon gas can be coarse-grained on scales relevant to photon-graviton mixing.  In practice, this allows us to discard all terms in the action that are odd in the background field strength, $\overline{ F}_{ab}$.  In this regime, the photon and graviton propagate independently, albeit with a modified dispersion relation induced by the ambient photon gas.  
To calculate the photon dispersion relation, we then simply extract the part of the effective action~\eqref{eq:EH2} that is quadratic in the photon fluctuation.   Note that while the energy of the propagating photon that we consider is, by construction, less than the temperature, the wavelength can still easily be much shorter than the typical scale of spacetime curvature induced by the photon gas.  The thermal background sources a conformally-flat FRW metric, which acts effectively as flat space for photon
propagation at leading order due to classical conformal invariance of electromagnetism in 4D; in any case, just as in \Sec{sec:Causality3}, a conformally-flat metric in any dimension reduces the question of causality to a special relativistic problem, since coordinate speeds and vielbein speeds coincide.

In 4D, the energy density $\rho$ and pressure $p$ are related by $p = \rho/3$, where 
$\rho = {\pi^2 } T^4/15$.  Using the fact that $\overline{T}^{ab} = \textrm{diag}(\rho,p,p,p)$ together with \Eq{eq:T}, we find the simple expression
\be 
\overline{F_{ab} F_{cd}} =\overline{\tilde F_{ab} \tilde F_{cd}} = \frac{\pi^2}{45}T^4(\delta_{ac}\delta_{bd} - \delta_{ad}\delta_{bc}),\label{eq:F2ab}
\ee 
where $\delta_{ab}$ is again the Kronecker delta function.  As in Eq.~\eqref{eq:dphiab}, Eq.~\eqref{eq:F2ab} breaks Lorentz invariance due to the existence of the preferred rest frame of the photon gas. Inputting this expression into the effective metric \eqref{eq:effmet4Weyl}, we find 
\be 
v = \frac{k_0}{|\vec k|} = 1- \frac{32 \pi^2}{45} (a_1'+a_2')T^4,
\label{eq:vFRW}
\ee
independent of the direction of propagation or polarization, where we have used that $\overline{C}_{abcd}=0$ because the background FRW metric is conformally flat.  In the limit that gravity is decoupled, our expression for the photon velocity  agrees with \Ref{Ravndal}, which considered a thermal photon background in flat space. Note, however, that our formula does not agree with \Ref{D&H}, which computed the photon velocity in a FRW universe but neglected to include the corrections coming from $(F_{\mu\nu} F^{\mu\nu})^2$ and $(F_{\mu\nu} \tilde F^{\mu\nu})^2$. 
In conclusion, we require that  
\be 
a_1' +a_2' \geq 0
\label{eq:causal4d}
\ee
 to forbid superluminal propagation within the photon gas.

The relationship between superluminality and causality violation is, however, quite subtle in curved spacetime.  A famous example is the seminal work of \Ref{D&H}, which computed the speed of photons near a Schwarzschild black hole, taking into account corrections from the gravitational Euler--Heisenberg Lagrangian obtained by integrating out the electron.  Curiously, the authors of \Ref{D&H} found that orbitally-traversing photons polarized in the radial direction propagate superluminally.  However, this superluminal propagation cannot be an authentic signal of causality violation since the theory is literally real-world electrodynamics.  While there is no universally-accepted resolution to this puzzle, it is important to note that an explicit CCC was not constructed in \Ref{D&H}.\footnote{It has been argued (see \Ref{Hollowood} and refs. therein) that the superluminality derived in \Ref{D&H} is harmless because causality is dictated by high-frequency photon modes that lie outside the regime of the photon-graviton effective theory.  However, this interpretation implies non-analyticity of the photon propagator and violation of the Kramers--Kronig dispersion relation.}  Despite the existence of local superluminal propagation, it is therefore clear that spacetime curvature can compensate for these effects in such a way that actual information flow remains causal.   This is a prime example of the fallacy of interpreting superluminality as a telltale sign of acausal signal propagation.

Our ideal goal is then to engineer a CCC in 4D that is analogous to the construction in \Sec{sec:Causality3}, consisting of two bubbles of thermal photon gas in relative motion.  However, since 4D gravity is dynamical, a non-vanishing Weyl tensor is induced in the vacuum region exterior to the photon gas.  As shown in \Eq{eq:compatibilityCCC}, if photons are only slightly superluminal, then a CCC requires a huge relative boost.  In turn, the curvature outside the bubbles will be large and thus important for the propagation of photons during their traversal between the bubbles.    Indeed, these metric effects will generally dominate over those induced by higher-dimension operators in the effective action.
In addition, at such large relative boosts, it is no longer a good approximation to treat the bubbles as independent because they back-react.  Of course, none of these effects arise in 3D, where the metric is locally flat in vacuum. 
Nonetheless, as we shall see, superluminal photon propagation can be linked to sharp pathologies via more elaborate constructions involving black holes.

In particular, consider a Schwarzschild black hole in the Hartle--Hawking vacuum \cite{HHvacuum}.  This describes a black hole in equilibrium with an exterior thermal bath, so the event horizon is static.\footnote{Without this stipulation, Hawking evaporation causes the event horizon to move faster than the tiny corrections to the speed of light that we consider here.} 
Outside the black hole, the energy-momentum tensor is approximately described by a thermal gas at  Hawking temperature $T$. For a sufficiently massive black hole, $T$ can easily lie below the cutoff of the photon-graviton effective theory.  The thermal background outside the black hole causes the speed of light to vary in accordance with our earlier discussion of FRW.  However, there is an additional subtlety here in that, unlike the FRW case, the Schwarzschild geometry is not conformally flat, so we must account for the coupling of propagating photons to the background Weyl tensor in \Eq{eq:effmet4Weyl}.  As shown in \Ref{D&H}, however, this contribution does not affect the speed of radially-propagating photons, so the Weyl component of the Schwarzschild metric can be ignored.   Another subtlety is that very close to the horizon, the Hartle--Hawking vacuum actually implies deviations from thermality \cite{Visser}.  Because of these differences, \Eq{eq:vFRW} does not, strictly speaking, apply; that is, the numerical details of the superluminality bound \eqref{eq:causal4d} may be somewhat different. In any case, these detailed near-horizon corrections affect our results quantitatively but not qualitatively.  

Consider the case in which the superluminality bound fails.  In this case, photons will traverse slightly outside of the light-cone defined by the spacetime metric, due to the ambient Hawking radiation.  Note that this setup differs crucially from that of \Ref{D&H}.  In particular, the authors of \Ref{D&H} did not consider the effects of Hawking radiation, so modifications to the photon speed arose solely from the non-vanishing Weyl tensor in the vacuum Schwarzschild spacetime.  As a result, \Ref{D&H} found that radially-propagating photons were luminal, so light cannot escape the event horizon.  On the other hand, in our construction radial photons are superluminal if the bound fails, because the Hawking radiation modifies the photon speed in all directions.   Consequently, a signal sent from inside the horizon can propagate radially to the outside in finite time as measured by an exterior observer.  This phenomenon is in tension with black hole complementarity \cite{Complementarity}, in which the exterior and interior regions are treated as separate but equivalent Hilbert spaces. That is, if one were able to send signals from behind the horizon of a black hole, then the usual challenges to unitarity that come from black hole information theory \cite{PreskillMirror} would no longer be so elegantly solved by complementarity.

Alternatively, one can interpret deviations from luminal photon propagation as a modification of the effective horizon of the black hole.  For example, take the case where \Eq{eq:causal4d} (or its near-horizon analogue) is violated and the photon is superluminal due to the ambient Hawking radiation.  The effective horizon tilts in the space-like direction, shifting to a radius smaller than the usual Schwarzschild radius.  
Because the effective horizon shrinks, Hawking-radiated photons are emitted at a higher temperature.   As the temperature increases, the velocity shift of the photon then increases, thus shrinking the effective horizon even more.  In principle, this suggests an instability in the position of the effective black hole horizon.  In contrast, if the bound is satisfied, then photon propagation is subluminal, the effective horizon grows, and Hawking-radiated photons exit at a lower temperature.  In this case, the ambient photon gas is colder and the photon speed moves closer to unity.  Hence, in this scenario the position of the effective horizon is stable.

\begin{figure}[t] 
\centerline{\includegraphics[width=0.59\columnwidth]{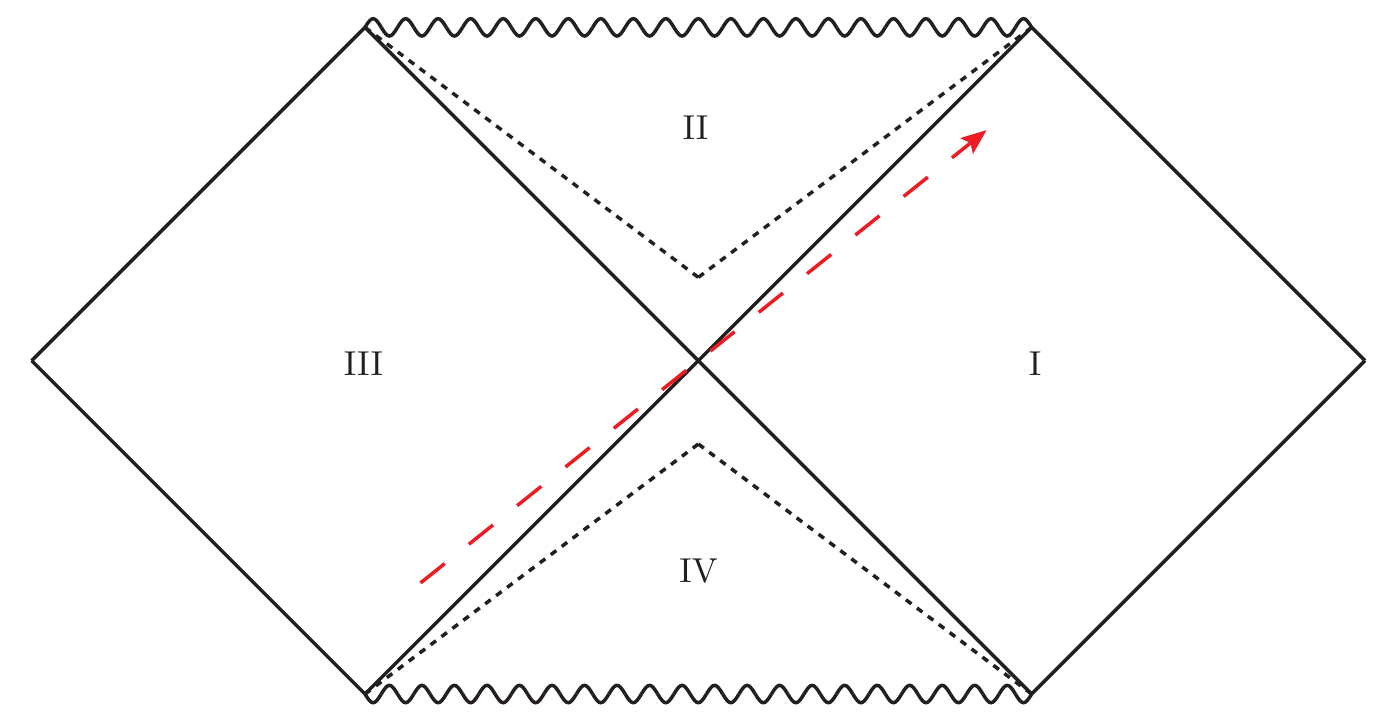}}
\caption{Conformal diagram for a maximally-extended Schwarzschild black hole. The effective horizon (dotted black) shrinks in a theory failing our superluminality bound. Superluminal photon propagation (red dashed arrow) allows observers in regions I and III to communicate.}\label{fig:kruskalBH}
\end{figure}

\begin{figure}[t] 
\centerline{
\includegraphics[
width=.95\columnwidth]{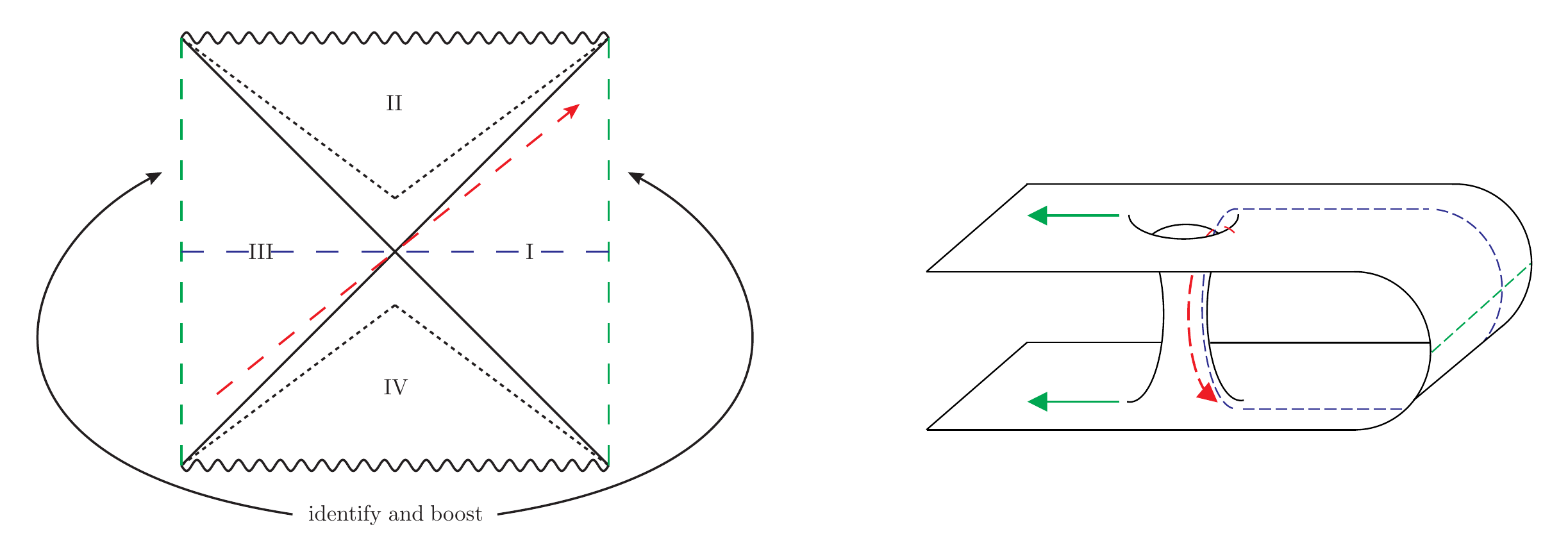}}\vspace*{-3mm}
\caption{Conformal diagram (left) and embedding diagram of a spacelike slice (right) of the maximally-extended Schwarzschild black hole, describing wormhole mouths in relative motion. In a theory with superluminal propagation, the effective horizon (dotted black) shrinks and the wormhole becomes traversable by a signal sent from region III to I (red dashed arrow).  The codimension-one surfaces (dashed green) at large spatial distance from the mouths are identified, albeit boosted relative to one another (green arrows).  Also shown is a particular tangent codimension-three spacelike surface (dashed blue).}\label{fig:4dCCC}
\end{figure}

Last of all, let us consider the maximally-extended Schwarzschild solution \cite{Kruskal}.
This background supports two asymptotically-flat spacetime regions, I and III, exterior to the two-sided black hole.  One interpretation of this spacetime is that it describes a wormhole linking two black hole mouths \cite{Causality}.  When the superluminality bound fails, the concomitant faster-than-light propagation enables observers in regions I and III to communicate by sending signals through region II,\footnote{As for the one-sided black hole, we require that both wormhole mouths have static event horizons, which can be achieved by putting each in equilibrium with a thermal bath enclosing the mouth.} as shown in \Fig{fig:kruskalBH}. Physically, this implies that the Einstein--Rosen bridge is traversable by photons and thus regions I and III are in causal contact.   In contrast with usual constructions of traversable wormholes, this setup does not require the existence of exotic matter and associated violations of the averaged null energy condition \cite{CTCs}. As discussed in \Ref{wormholes}, if the wormhole mouths are in relative motion, it is possible to construct a CCC, in this case not traversable by matter following timelike or null trajectories, but rather by the superluminal photons that result from violation of the near-horizon version of \Eq{eq:causal4d}, which is equally destructive to causality. See \Fig{fig:4dCCC} for an illustration of this setup.

Note, however, that a wormhole can only support a true causal paradox if there is a boost between the wormhole mouths. In essence, the CCC construction is similar to that of \Sec{sec:Causality3}, with the difference being that here we consider signals sent through the wormhole between two observers, one located just outside of each wormhole mouth. 
In particular, if the mouths are in relative motion at velocity $u$, then \Eq{eq:compatibilityCCC} must be satisfied, where $v_{\rm avg}$ is the effective speed at which a light signal appears to propagate between the mouths as seen in the exterior spacetime, {\it i.e.}, the speed of information propagation as measured by external observers located near each wormhole mouth.   For $v_{\rm avg}$ only slightly superluminal, an enormous boost is required, inducing large back-reaction on the metric and invalidating our starting background. However, the wormhole mouths can be taken to be parametrically far apart; since the time the signal takes to go through the wormhole throat is independent of the distance between the mouths, $v_{\rm avg}$ can be made arbitrarily superluminal, overcoming any gravitational redshift effect in the exterior spacetime. With $v_{\rm avg}$ parametrically large, the required boost $u$ can be very small, yielding negligible back-reaction and gravitational radiation while still allowing for the formation of a CCC. 

From the perspective of AdS/CFT \cite{AdSCFT,AdSCFT2}, signal propagation through a traversable wormhole is puzzling and likely pathological \cite{graviton3}.  As observed in \Ref{Sundrum}, traversable wormholes correspond to non-local dynamics in the dual CFT.   More concretely, our particular setup can be embedded in the construction of \Ref{VanRaamsdonk}: a  maximally-extended Schwarzschild black hole geometry in asymptotically-AdS spacetime, dual to  two entangled non-interacting CFTs on a sphere. In this geometry, the ability to send signals between regions I and III is dual to non-unitary evolution of the CFT, thus disrupting the canonical notions of entanglement entropy between the two CFTs \cite{RT,Stefan}. Moreover, in light of the ER=EPR conjecture \cite{EREPR}, communication between mouths of an Einstein--Rosen bridge is dual to pathological information transfer via  entanglement.   While this scenario is deserving of a more thorough analysis, it lies beyond the scope of the present work.

We have outlined a variety of causal and quantum gravitational pathologies that suggest that a superluminality bound like \Eq{eq:causal4d} is a requirement of any consistent low-energy effective theory.   Assuming we are permitted to locate regions I and III of the extended Schwarzschild solution within the same asymptotic spacetime, then corrections to photon propagation that violate the superluminality bound transform the Einstein--Rosen bridge into a traversable wormhole and a CCC can be formed.

\section{Summary and Future Directions}\label{sec:conclusions}

In this paper, we have derived infrared consistency conditions on the photon-graviton effective action in \Eq{eq:EH} in 3D and 4D.  These bounds are deduced from considerations of analyticity of light-by-light scattering, unitarity of the ultraviolet completion, and superluminality of photon fluctuations in non-trivial backgrounds.  The 3D setup is a convenient starting point, where gravity is non-dynamical but still has a physical effect on photon-photon interactions.  In 4D, many of the arguments are complicated (or, in the case of analyticity, even obstructed) by dynamical gravity.  Our bounds on the photon-graviton effective action are summarized in Eqs.~\eqref{eq:introbound3d}, \eqref{eq:introbound4d2}, and \eqref{eq:introbound4d1} in \Sec{sec:intro}.  We then specialize to the case where electromagnetic corrections to the effective action come from a  particle of charge-to-mass ratio $z=q/m$.  Our infrared consistency conditions are then a constraint on a combination of $z$ and coefficients parameterizing unspecified gravitational corrections, as shown in Eqs.~\eqref{eq:3Dcondition}, \eqref{eq:UnitarityBound4D}, and \eqref{eq:CausalityBound4D} and in \Fig{fig:bound4d}.  

The present work leaves a number of interesting avenues for future research. For example, as noted in \Ref{HierarchyPaper}, the WGC is not sharply defined in a theory with a Higgsed Abelian force carrier.  In particular, in the Higgs phase, states of different charge can mix, so $q$ and $m$ are non-commuting operators, thus making the WGC ill-defined.  Furthermore, the original justification of the WGC---that is, the pathology of exactly stable extremal black holes---is murky in the Higgs phase since a charged black hole can shed charge associated with a massive $U(1)$ and subsequently decay.  On the other hand, the photon-graviton effective action is still well-defined irrespective of whether the photon is massive or massless.  As a result, it is especially interesting to consider infrared consistency conditions in the presence of a non-zero photon mass. 
For a Proca theory, we can simply add a physical mass.  A more interesting case would be to introduce dynamical gauge symmetry breaking with a physical Higgs field.

Another direction for future work relates to the more complicated scenario of multiple Abelian forces.  As shown in \Ref{HierarchyPaper}, it is straightforward to apply the logic of extremal black hole decay to theories with multiple forces and charged particles.  The generalization of the WGC then becomes a simple geometric condition on the vectors describing the charge-to-mass ratios of particles in the theory.  This generalization demands a more stringent constraint than \Eq{eq:WGC} applied to each charge axis.  Given this understanding, it would be interesting to see if similar geometric constraints arise from studying the low-energy effective action describing multiple photons interacting with the graviton.  In principle, such an action will have many more free parameters than \Eq{eq:EH}, but likewise many more constraints coming from analyticity, unitarity, and causality.

Last of all, we have not pursued possible constraints on the photon-graviton action from thermodynamic considerations.  As discussed in \Ref{PerpetuumMobile}, variations in the speed of light can allow for violation of the second law of thermodynamics when considering Hawking radiation in a black hole background.  Since the speed of photon propagation is modified by higher-dimension operators, it may be possible to derive additional substantive constraints from thermodynamic reasoning.

The boundary between the landscape of healthy ultraviolet-completable theories and the swampland of pathological effective theories offers a promising arena for new physics insights. 
As we have shown, the particular criterion asserted by the WGC may be studied from purely low-energy reasoning given the non-trivial requirements of infrared consistency.  In particular, we have determined regions in the effective theory that are forbidden by
 violations of analyticity, unitarity, and causality.   Rescuing the forbidden regions of parameter space would require loopholes in all three arguments, or alternatively, reasons to countenance all of these pathologies.

\begin{center} 
 {\bf Acknowledgments}
 \end{center}
 \noindent 
We thank Allan Adams, Nima Arkani-Hamed, Brando Bellazzini, Sean Carroll, Stanley Deser, Tim Hollowood, Stefan Leichenauer, Alberto Nicolis, Rafael Porto, and Ira Rothstein for useful discussions and comments. C.C.~is supported by a Sloan Research Fellowship and a DOE Early Career Award under Grant No.~DE-SC0010255.  G.N.R.~is supported by a Hertz Graduate Fellowship and a NSF Graduate Research Fellowship under Grant No.~DGE-1144469.

\appendix
\section{Appendix}
\label{appendix}
\setcounter{equation}{0}
\numberwithin{equation}{section}
In \Eq{eq:Pi4}, we introduced a spectral representation for the field $\chi^{(4)}_{\mu\nu\rho\sigma}$.  We now show that the tensor numerator of this spectral representation, $\Pi_{\mu\nu\rho\sigma\alpha\beta\gamma\delta}$, is highly constrained.  To begin, note that $\chi^{(4)}_{\mu\nu\rho\sigma}$ does not correspond to a canonical spin-4 state, which is traditionally represented by a four-index, fully symmetric tensor \cite{Weinberg1,Weinberg3,Fronsdal,arbitraryspin}.  Like the Riemann tensor, $\chi^{(4)}_{\mu\nu\rho\sigma}$ is instead antisymmetric in its first and second pairs of indices separately and symmetric on the exchange of these pairs. 
The projection operator $\Pi_{\mu\nu\rho\sigma\alpha\beta\gamma\delta}$ inherits these index symmetry properties and tracelessness, and is furthermore symmetric on the interchange of the entire first and second sets of four indices. To determine $\Pi_{\mu\nu\rho\sigma\alpha\beta\gamma\delta}$, we start with an ansatz tensor that is an arbitrary function of $\eta_{\mu\nu}$ and $k_\mu$.  Imposing the transverse condition $k^\mu \Pi_{\mu\nu\rho\sigma\alpha\beta\gamma\delta} = 0$ on-shell, it is straightforward to show that $\Pi_{\mu\nu\rho\sigma\alpha\beta\gamma\delta}$ is necessarily a function of the projection operator $\Pi_{\mu\nu}$ in \Eq{eq:projection}.  Altogether, these restrictions only allow for two possible tensor structures: 
\begin{align}
&\hspace{9.1mm}\Pi_{\mu\rho}\Pi_{\nu\beta}\Pi_{\sigma\delta}\Pi_{\alpha\gamma}+\Pi_{\nu\sigma}\Pi_{\mu\beta}\Pi_{\rho\delta}\Pi_{\alpha\gamma}+\Pi_{\mu\rho}\Pi_{\nu\alpha}\Pi_{\sigma\gamma}\Pi_{\beta\delta}+\Pi_{\nu\sigma}\Pi_{\mu\alpha}\Pi_{\rho\gamma}\Pi_{\beta\delta}\nn\\
 & \hspace{4mm}+\Pi_{\nu\rho}\Pi_{\mu\gamma}\Pi_{\sigma\beta}\Pi_{\alpha\delta}+\Pi_{\mu\sigma}\Pi_{\nu\gamma}\Pi_{\rho\beta}\Pi_{\alpha\delta}+\Pi_{\nu\rho}\Pi_{\mu\delta}\Pi_{\sigma\alpha}\Pi_{\beta\gamma}+\Pi_{\mu\sigma}\Pi_{\nu\delta}\Pi_{\rho\alpha}\Pi_{\beta\gamma}\nn\\
 & \hspace{4mm}-\Pi_{\nu\rho}\Pi_{\mu\beta}\Pi_{\sigma\delta}\Pi_{\alpha\gamma}-\Pi_{\mu\sigma}\Pi_{\nu\beta}\Pi_{\rho\delta}\Pi_{\alpha\gamma}-\Pi_{\nu\rho}\Pi_{\mu\alpha}\Pi_{\sigma\gamma}\Pi_{\beta\delta}-\Pi_{\mu\sigma}\Pi_{\nu\alpha}\Pi_{\rho\gamma}\Pi_{\beta\delta}\nn\\
 & \hspace{4mm}-\Pi_{\mu\rho}\Pi_{\nu\alpha}\Pi_{\sigma\delta}\Pi_{\beta\gamma}-\Pi_{\nu\sigma}\Pi_{\mu\alpha}\Pi_{\rho\delta}\Pi_{\beta\gamma}-\Pi_{\mu\rho}\Pi_{\nu\beta}\Pi_{\sigma\gamma}\Pi_{\alpha\delta}-\Pi_{\nu\sigma}\Pi_{\mu\beta}\Pi_{\rho\gamma}\Pi_{\alpha\delta}\nn\\
 & \hspace{4mm}+\Pi_{\nu\rho}\Pi_{\mu\alpha}\Pi_{\sigma\delta}\Pi_{\beta\gamma}+\Pi_{\mu\sigma}\Pi_{\nu\alpha}\Pi_{\rho\delta}\Pi_{\beta\gamma}+\Pi_{\nu\rho}\Pi_{\mu\beta}\Pi_{\sigma\gamma}\Pi_{\alpha\delta}+\Pi_{\mu\sigma}\Pi_{\nu\beta}\Pi_{\rho\gamma}\Pi_{\alpha\delta}\nn\\
 & \hspace{4mm}+\Pi_{\mu\rho}\Pi_{\nu\delta}\Pi_{\sigma\beta}\Pi_{\alpha\gamma}+\Pi_{\nu\sigma}\Pi_{\mu\delta}\Pi_{\rho\beta}\Pi_{\alpha\gamma}+\Pi_{\mu\rho}\Pi_{\nu\gamma}\Pi_{\sigma\alpha}\Pi_{\beta\delta}+\Pi_{\nu\sigma}\Pi_{\mu\gamma}\Pi_{\rho\alpha}\Pi_{\beta\delta}\nn\nn\\
 & \hspace{4mm}-\Pi_{\nu\rho}\Pi_{\mu\delta}\Pi_{\sigma\beta}\Pi_{\alpha\gamma}-\Pi_{\mu\sigma}\Pi_{\nu\delta}\Pi_{\rho\beta}\Pi_{\alpha\gamma}-\Pi_{\nu\rho}\Pi_{\mu\gamma}\Pi_{\sigma\alpha}\Pi_{\beta\delta}-\Pi_{\mu\sigma}\Pi_{\nu\gamma}\Pi_{\rho\alpha}\Pi_{\beta\delta}\nn\\
 & \hspace{4mm}-\Pi_{\mu\rho}\Pi_{\nu\gamma}\Pi_{\sigma\beta}\Pi_{\alpha\delta}-\Pi_{\nu\sigma}\Pi_{\mu\gamma}\Pi_{\rho\beta}\Pi_{\alpha\delta}-\Pi_{\mu\rho}\Pi_{\nu\delta}\Pi_{\sigma\alpha}\Pi_{\beta\gamma}-\Pi_{\nu\sigma}\Pi_{\mu\delta}\Pi_{\rho\alpha}\Pi_{\beta\gamma}\nn\\
 & \hspace{4mm}-\Pi_{\mu\alpha}\Pi_{\nu\beta}\Pi_{\rho\gamma}\Pi_{\sigma\delta}+\Pi_{\mu\beta}\Pi_{\nu\alpha}\Pi_{\rho\gamma}\Pi_{\sigma\delta}+\Pi_{\mu\alpha}\Pi_{\nu\beta}\Pi_{\rho\delta}\Pi_{\sigma\gamma}-\Pi_{\mu\beta}\Pi_{\nu\alpha}\Pi_{\rho\delta}\Pi_{\sigma\gamma}\nn\\
 & \hspace{4mm}-\Pi_{\mu\gamma}\Pi_{\nu\delta}\Pi_{\rho\alpha}\Pi_{\sigma\beta}+\Pi_{\mu\delta}\Pi_{\nu\gamma}\Pi_{\rho\alpha}\Pi_{\sigma\beta}+\Pi_{\mu\gamma}\Pi_{\nu\delta}\Pi_{\rho\beta}\Pi_{\sigma\alpha}-\Pi_{\mu\delta}\Pi_{\nu\gamma}\Pi_{\rho\beta}\Pi_{\sigma\alpha}\nn\\
 & \hspace{4mm}+2(\Pi_{\mu\rho}\Pi_{\nu\sigma}\Pi_{\alpha\delta}\Pi_{\beta\gamma}-\Pi_{\mu\rho}\Pi_{\nu\sigma}\Pi_{\alpha\gamma}\Pi_{\beta\delta}-\Pi_{\mu\sigma}\Pi_{\nu\rho}\Pi_{\alpha\delta}\Pi_{\beta\gamma}+\Pi_{\mu\sigma}\Pi_{\nu\rho}\Pi_{\alpha\gamma}\Pi_{\beta\delta})
 \label{eq:bigPi1}\\
 \intertext{and}
 &\hspace{9.1mm}\Pi_{\mu\alpha}\Pi_{\nu\beta}\Pi_{\rho\gamma}\Pi_{\sigma\delta}-\Pi_{\mu\beta}\Pi_{\nu\alpha}\Pi_{\rho\gamma}\Pi_{\sigma\delta}-\Pi_{\mu\alpha}\Pi_{\nu\beta}\Pi_{\rho\delta}\Pi_{\sigma\gamma}+\Pi_{\mu\beta}\Pi_{\nu\alpha}\Pi_{\rho\delta}\Pi_{\sigma\gamma}\nn\\
 & \hspace{4mm}+\Pi_{\mu\gamma}\Pi_{\nu\delta}\Pi_{\rho\alpha}\Pi_{\sigma\beta}-\Pi_{\mu\delta}\Pi_{\nu\gamma}\Pi_{\rho\alpha}\Pi_{\sigma\beta}-\Pi_{\mu\gamma}\Pi_{\nu\delta}\Pi_{\rho\beta}\Pi_{\sigma\alpha}+\Pi_{\mu\delta}\Pi_{\nu\gamma}\Pi_{\rho\beta}\Pi_{\sigma\alpha}\nn\\
 & \hspace{4mm}+\Pi_{\mu\alpha}\Pi_{\nu\gamma}\Pi_{\rho\delta}\Pi_{\sigma\beta}-\Pi_{\mu\gamma}\Pi_{\nu\alpha}\Pi_{\rho\delta}\Pi_{\sigma\beta}-\Pi_{\mu\alpha}\Pi_{\nu\gamma}\Pi_{\rho\beta}\Pi_{\sigma\delta}+\Pi_{\mu\gamma}\Pi_{\nu\alpha}\Pi_{\rho\beta}\Pi_{\sigma\delta}\nn\\
 & \hspace{4mm}+\Pi_{\mu\alpha}\Pi_{\nu\delta}\Pi_{\rho\beta}\Pi_{\sigma\gamma}-\Pi_{\mu\delta}\Pi_{\nu\alpha}\Pi_{\rho\beta}\Pi_{\sigma\gamma}-\Pi_{\mu\alpha}\Pi_{\nu\delta}\Pi_{\rho\gamma}\Pi_{\sigma\beta}+\Pi_{\mu\delta}\Pi_{\nu\alpha}\Pi_{\rho\gamma}\Pi_{\sigma\beta}\nn\\
 & \hspace{4mm}+\Pi_{\mu\gamma}\Pi_{\nu\beta}\Pi_{\rho\delta}\Pi_{\sigma\alpha}-\Pi_{\mu\beta}\Pi_{\nu\gamma}\Pi_{\rho\delta}\Pi_{\sigma\alpha}-\Pi_{\mu\gamma}\Pi_{\nu\beta}\Pi_{\rho\alpha}\Pi_{\sigma\delta}+\Pi_{\mu\beta}\Pi_{\nu\gamma}\Pi_{\rho\alpha}\Pi_{\sigma\delta}\nn\\
 & \hspace{4mm}+\Pi_{\mu\delta}\Pi_{\nu\beta}\Pi_{\rho\alpha}\Pi_{\sigma\gamma}-\Pi_{\mu\beta}\Pi_{\nu\delta}\Pi_{\rho\alpha}\Pi_{\sigma\gamma}-\Pi_{\mu\delta}\Pi_{\nu\beta}\Pi_{\rho\gamma}\Pi_{\sigma\alpha}+\Pi_{\mu\beta}\Pi_{\nu\delta}\Pi_{\rho\gamma}\Pi_{\sigma\alpha}  
 .\label{eq:bigPi2}
\end{align}
Consequently, $\Pi_{\mu\nu\rho\sigma\alpha\beta\gamma\delta}$ must be an arbitrary linear combination of these two tensors.  
As noted in the body of the text, however, the forms of these tensors imply that $\Pi_{\mu\nu\rho\sigma}^{\;\;\;\;\;\;\;\;\; \mu\nu\rho\sigma} =0$, which cannot be equal to a sum over polarization tensors and is thus in violation of unitarity.

\bibliography{WGCbib}
\bibliographystyle{utphys}
\end{document}